\documentclass[twocolumn]{aastex62}
\usepackage{natbib, chngcntr}
\usepackage{stmaryrd}
\usepackage{enumitem}
\renewcommand{\tabcolsep}{3.2pt}
\usepackage{mathtools, amsmath}
\usepackage{xcolor}
\usepackage{placeins}
\usepackage{graphicx}
\usepackage{array}
\usepackage{booktabs}

\begin{document}

\title{Analyzing Atmospheric Temperature Profiles and Spectra of M dwarf Rocky Planets}
\author{Matej Malik}
\affiliation{Department of Astronomy, University of Maryland, College Park, MD 20742, USA}

\author{Eliza M.-R. Kempton}
\affiliation{Department of Astronomy, University of Maryland, College Park, MD 20742, USA}
\affil{Department of Physics, Grinnell College, 1116 8th Avenue, Grinnell, IA 50112, USA}

\author{Daniel D. B. Koll}
\affiliation{Department of Earth, Atmospheric, and Planetary Sciences, Massachusetts Institute of Technology, Cambridge, MA 02139, USA}

\author{Megan Mansfield}
\affiliation{Department of Geophysical Sciences, University of Chicago, Chicago, IL 60637, USA}

\author{Jacob L. Bean}
\affiliation{Department of Astronomy \& Astrophysics, University of Chicago, Chicago, IL 60637, USA}

\author{Edwin Kite}
\affiliation{Department of Geophysical Sciences, University of Chicago, Chicago, IL 60637, USA}

\correspondingauthor{Matej Malik}
\email{malik@umd.edu}

\shorttitle{Atmospheric Characteristics of Planets around M Stars}
\shortauthors{M. Malik, E. M.-R. Kempton, D. D. B. Koll, et al.}

\begin{abstract}

The James Webb Space Telescope ({\it JWST}) will open up the possibility of comprehensively measuring the thermal emission spectra of rocky exoplanets orbiting M dwarfs to characterize their atmospheres. In preparation for this opportunity, we present model atmospheres for three M-dwarf planets particularly amenable to secondary eclipse spectroscopy --- TRAPPIST-1b, GJ 1132b, and LHS 3844b. Using three limiting cases of candidate atmospheric compositions (pure H$_2$O, pure CO$_2$ and solar abundances) we calculate temperature-pressure profiles and emission spectra in radiative-convective equilibrium, including the effects of a solid surface. We find that the atmospheric radiative transfer is significantly influenced by the cool M-star irradiation; H$_2$O and CO$_2$ absorption bands in the near-infrared are strong enough to absorb a sizeable fraction of the incoming stellar light at low pressures, which leads to temperature inversions in the upper atmosphere. The non-gray band structure of gaseous opacities in the infrared is hereby an important factor. Opacity windows are muted at higher atmospheric temperatures, so we expect temperature inversions to be common only for sufficiently cool planets. We also find that pure CO$_2$ atmospheres exhibit lower overall temperatures and stronger reflection spectra compared to models of the other compositions. We estimate that for GJ 1132b and LHS 3844b we should be able to distinguish between different atmospheric compositions with {\it JWST}. The emission lines from the predicted temperature inversions are currently hard to measure, but high resolution spectroscopy with future Extremely Large Telescopes may be able to detect them.

\end{abstract}


\section{Introduction}
\label{sec:intro}

With the {\it Transiting Exoplanet Survey Satellite} ({\it TESS}) conducting an all-sky survey, we have now the opportunity to detect a windfall of planets orbiting bright nearby M-type stars \citep{ricker15}. Small rocky planets are the prize of any space-based transit survey, and such planets are known to be commonplace around M dwarfs \citep{dressing15, mulders15, gaidos16}. Also, the geometric probability of finding transiting planets of a given temperature is greater for smaller stars than stars of other, larger types. In addition, M dwarfs are intrinsically less luminous so that the relative spectroscopic signal of a transiting planet is larger. The shift to cooler stars also helps our efforts on searching for habitable planets. Whereas habitable planets around G stars reside on wide orbits with periods $\sim 100$ days, habitable zones around M stars are located much closer in, with periods of $\sim$ few days, enhancing the crucial repeatability of observations \citep{kopparapu13, kopparapu16}.

Although terrestrial, potentially habitable planets are among the most fascinating candidates for characterization surveys, they are yet mostly out of reach with current instrumentation \citep{cowan15}. Thus, a more pragmatic approach is to focus on warm (i.e.\ hotter than habitable) rocky super-Earths. Such planets may have experienced a similar evolution as their Earth-analogue counterparts and thus may possess similar atmospheric and interior properties, yet, due to their larger size and higher temperatures, provide the advantage of a better signal.

Still, even in this regime our observational data are limited, providing us with few constraints on the planetary and atmospheric properties of those objects. One way to improve our understanding is to build self-consistent atmospheric models based on first principles. Not only do these models strengthen our physical intuition, they also allow us to make predictions about the planets' spectral appearance, serving as guidance for future observations.

Modeling atmospheres of rocky planets is not a new endeavor, after all there are such objects in the Solar System. However, only a few other works have specifically studied the radiative transfer and thermal emission spectra of the near-term observable exoplanets around M stars. Recent assessments of the observability of the TRAPPIST-1 planets, GJ 1132b and LHS 1140b with {\it JWST} \citep{barstow16, morley17, batalha18} focused on mapping out the optimal observation strategies and the connected uncertainties and relied on simplified radiative transfer schemes. On the other side of the spectrum are the habitability studies of \citet{wolf17, lincowski18, meadows18}, which explore the potential habitability of the temperate planets in the TRAPPIST-1 system and Proxima Centauri b. Using sophisticated atmospheric models, like three-dimensional global climate models, based on our Solar System knowledge, their meticulous examination is optimized to study individual planets, but may lack flexibility when interested in a wider parameter regime of known and yet to be discovered super-Earths and terrestrial planets. 

In this work, we choose an intermediate approach, using a radiative-convective model to self-consistently assess the temperature profiles and corresponding thermal emission and reflection spectra of hot rocky exoplanets, incorporating the radiative balance with a solid surface at the base of the atmosphere. We focus on the M dwarf planets TRAPPIST-1b, GJ 1132b and LHS 3844b as they are currently the best targets for emission spectroscopy of rocky, terrestrial planets and representative of future {\it TESS} discoveries amenable for follow-up spectroscopic {\it JWST} observations \citep{kempton18}. 

The main advantage of thermal emission spectroscopy over the more widely used transmission spectroscopy is that it suffers less from signal loss if the atmosphere is cloudy. Furthermore, the transmission signal is diminished for atmospheres with a high mean molecular weight, which is not the case for the thermal emission signal (e.g., \citealt{miller-ricci09}). Thermal emission spectra also provide a direct probe of the atmospheric structure (i.e.\ vertical temperature gradients). Finally, \citet{morley17} found that hotter terrestrial planets are easier to characterize by emission than by transmission spectroscopy. However, due to their choice of parametrized vertical temperature profiles they may have overpredicted the emission signal strength.

In terms of the atmospheric composition, we construct numerical models of pure water and pure CO$_2$ atmospheres and, in order to have covered another cornerstone of possible atmospheric conditions, add the case of a solar composition hydrogen-dominated, low-mean-molecular-weight atmosphere, even though the explored planets are unlikely to retain such an atmosphere over their lifetime. The notion seems reasonable that future exoplanet discoveries with {\it TESS} are likely to exhibit similar planetary and atmospheric properties to the span of models explored in this work.

Our knowledge about the specific atmospheric properties of the three test planets in this study is scarce. The mean density of TRAPPIST-1b suggests a volatile-rich envelope with of 10 - 10$^4$ bars pressure \citep{grimm18}, even though the planet probably already lost a large fraction of its initial mass due to photo-evaporation \citep{bolmont17,dong18}. Spectroscopic analyses have so far indicated a feature-less transmission spectrum, consistent with a high mean-molecular weight composition, as given by a water-dominated or Venus-like atmosphere, and have ruled out a cloud-free, hydrogen-dominated atmosphere \citep{dewit16,zhang18}. 

For GJ 1132b, even though planetary evolution calculations favor a tenuous atmosphere, a thick steam atmosphere is possible \citep{schaefer16} and a high mean molecular weight atmosphere is consistent with observations \citep{southworth17,diamond18}. 

The reported photometric radius of LHS 3844b of 1.3 $R_{\rm Earth}$ at the {\it TESS} and {\it MEarth} bandpasses \citep{vanderspek19} lies below the radius gap presented in \citet{fulton17} and the photo-evaporation valley \citep{owen13,owen17}, suggesting that the planet does not possess a primordial hydrogen envelope. This is consistent with the very recent {\it Spitzer} observations of LHS 3844b by \citet{kreidberg19}, whose analysis indicates a bare rock planet with no extended gaseous envelope. Even if this particular planet does not possess a thick atmosphere, we take it as an example for future discoveries of hotter rocky planets, which with different evolution pathways could still possess primordial or secondary envelopes.

In general, there is an ongoing debate on the extent to which super-Earths orbiting M stars are expected to possess substantial atmospheres at all. Young M stars exhibit a comparatively strong XUV flux, which depending on the planet's atmospheric composition, may lead to loss of the envelope due to photo-evaporation \citep{tian09, lopez12, owen13, jin14, chen16}. Another possibility is that some planets form at the end of the accretion disk phase and undergo most of their evolution after the gas disk dispersal. These planets would never have extended envelopes to begin with \citep{lopez18}. Thus, our model atmospheres are constructed under the assumption that the explored planets were able to retain their atmospheres or outgas secondary ones \citep{elkins08, kite09, rogers10}.

Our methodology is presented in detail in Sect.~\ref{sec:meth}. In these models we look for unique characteristics, which can help constrain the properties of the planets' atmospheres (see Sect.~\ref{sec:planet_grid}). We find that the low stellar temperature leads to unexpected consequences on the radiative transfer in the planet atmospheres. In Sect.~\ref{sec:inv}, we explore the dependency of high-altitude temperature inversions on the planetary and stellar parameters and offer an explanation for why they only emerge in the cooler temperature regime. We discuss the observability of the modeled atmospheres and inversions in Sect.~\ref{sec:detect}, comment on our assumptions and chosen parameters in Sect.~\ref{sec:param} and compare with other works in Sect.~\ref{sec:comp}.  We conclude in Sect.~\ref{sec:concl}.

\section{Method}
\label{sec:meth}

\begingroup
\renewcommand{\arraystretch}{1.3}
\begin{deluxetable}{l r r r r r r}
\tablecaption{
    Planetary system parameters used in this study.
    \label{tab:parameters}
    }
\tablecolumns{7}
\tablehead{
Planet & $R_{\rm pl}$ & $\log_{10} g$ & a & $R_{\rm star}$ & $T_{\rm star}$ & $T_{\rm eq}$\tablenotemark{b} \\
& ($R_{\Earth}$) & ($\log_{10}$ cm s$^{-2}$) & (AU) & ($R_{\Sun}$) & (K) & (K)
}
\startdata
TRAPPIST-1b\tablenotemark{1} & 1.12 & 2.9 & 0.0115 & 0.121 & 2511 & 393 \\
GJ 1132b\tablenotemark{2} & 1.16 & 3.068 & 0.0153 & 0.207 & 3270 & 580 \\
LHS 3844b\tablenotemark{3} & 1.32 & 3.112\tablenotemark{a} & 0.00622 & 0.189 & 3036 & 806 \\
\enddata
\tablenotemark{1}\cite{delrez18, grimm18}
\tablenotemark{2}\cite{berta15, bonfils18}
\tablenotemark{3}\cite{vanderspek19}
\tablenotetext{a}{Estimated with 2.3 $M_{\Earth}$, based on \citet{chen17}.}
\tablenotetext{b}{Assuming a zero Bond albedo.}
\end{deluxetable}
\endgroup

\begin{figure}
\begin{center}
\begin{minipage}{0.48\textwidth}
\includegraphics[width=\textwidth]{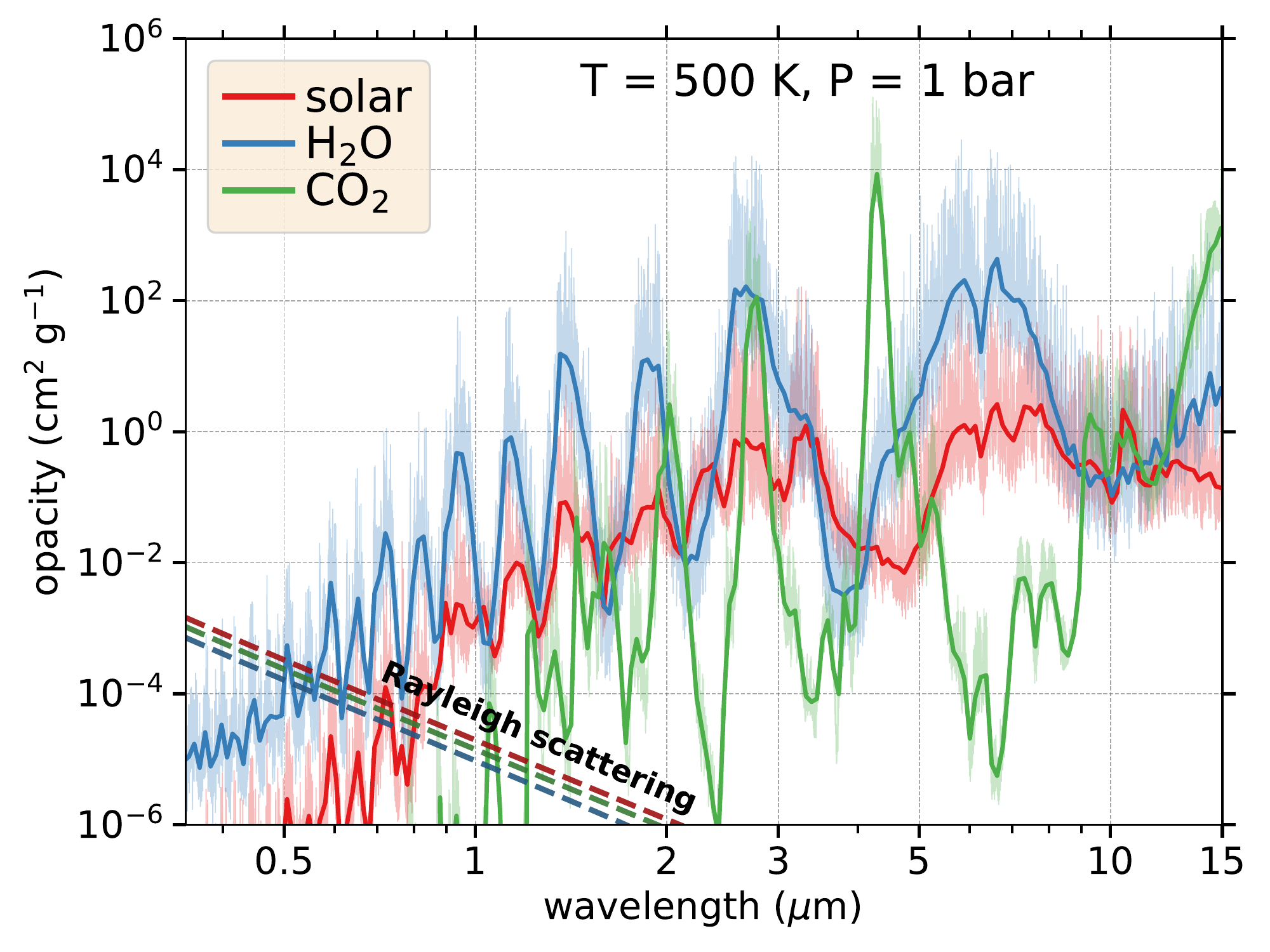}
\end{minipage}
\vspace{-0.3cm}
\caption{The opacity versus wavelength for the three atmospheric models used in this work, 100\% H$_2$O, 100\% CO$_2$, and solar abundances. The opacity function for the solar model is weighted with the mass mixing ratio of the respective constituents, obtained from equilibrium chemistry. Shown is an excerpt of the opacity tables for one temperature and pressure (500 K and 1 bar), including all the opacity sources for each model. Overplotted (dashed lines) is the contribution of the Rayleigh scattering cross-sections for the three models.}
\label{fig:opacities}
\end{center}
\end{figure}

The planetary parameters of TRAPPIST-1b, GJ 1132b, LHS 3844b are given in Table \ref{tab:parameters}. We use the open-source radiative transfer code \texttt{HELIOS}\footnote{\url{https://github.com/exoclime/HELIOS}} \citep{malik17, malik19} to calculate the temperature structures of these three planets in radiative-convective equilibrium and the corresponding synthetic emission and reflection spectra. We consider cloud-free atmospheres including Rayleigh scattering. The three simulated atmospheric models consist of 100\% H$_2$O, 100\% CO$_2$ and solar composition in chemical equilibrium. 

The solar model includes opacities calculated with HELIOS-K \citep{grimm15} from the following line lists: H$_2$O \citep{barber06}, CO$_2$ \citep{rothman10}, CO \citep{li15}, CH$_4$ \citep{yurchenko14}, NH$_3$ \citep{yurchenko11}, HCN \citep{harris06}, C$_2$H$_2$ \citep{gordon17}, PH$_3$ \citep{sousasilva15}, H$_2$S \citep{azzam16}, Na and K \citep{kurucz11, burrows00, burrows03}, and collision-induced absorption (CIA) of H$_2$-H$_2$ and H$_2$-He \citep{richard12}. The molecular lines are treated as Voigt profiles with a far-wing cut-off at 100 cm$^{-1}$ from line center. Pressure broadening is considered using the default parameters given in the ExoMol online database for H$_2$O, CO, CH4, NH$_3$, HCN, PH$_3$ and H$_2$S, and using the self-broadening prescription from the HITEMP/HITRAN database for CO$_2$ and C$_2$H$_2$. The treatment of the alkali opacities is described in detail in Appendix A of \citet{malik19}. For lower temperatures, Na and K are assumed to be removed from the gas phase by the formation of Na$_2$S and KCl condensates, respectively. We further include Rayleigh scattering cross-sections of H$_2$ and He \citep{cox00, lee04, sneep05, thalman14}. The chemical equilibrium mixing ratios are calculated with \texttt{FastChem} \citep{stock18}, taking the solar elemental abundances as reported in Table~1 of \citet{asplund09}. \texttt{FastChem} is tested between 100 K and 6000 K to accurately calculate the thermochemical equilibrium mixing ratios from a chemical network of 550 gas-phase species, including ions. 

The H$_2$O model includes H$_2$O line opacities and H$_2$O Rayleigh scattering cross-sections \citep{cox00, wagner08}, and the CO$_2$ model includes CO$_2$ line opacities, CO$_2$ scattering cross-sections \citep{cox00, sneep05, thalman14}, and CO$_2$-CO$_2$ CIA \citep{richard12}. The opacity functions of the three atmospheric models used in this work are shown in Fig.~\ref{fig:opacities}.

HELIOS utilizes a hemispheric two-stream method with improved scattering accuracy \citep{heng18} and convective adjustment, for which we set the adiabatic coefficient $\kappa \equiv (d \ln T / d \ln P)_S$, with the temperature  $T$, the pressure $P$, and the entropy $S$, to 2/7 for the solar case, and to 1/4 for the H$_2$O and CO$_2$ cases. For the temperature iteration we use 300 spectral bins between 0.33 $\mu$m and 10$^5$ $\mu$m with 20 Gaussian points in each bin, assuming a $k$-correlation. The emission spectra are generated at a resolution of $R=3000$ utilizing opacity sampling.

In order to obtain pronounced spectral features we simulate sufficiently thick atmospheres with a surface pressure of 10 bar. We set the heat redistribution factor $f$ \citep{burrows08, spiegel10} to $0.25$, indicating maximally efficient heat circulation, based on \citet{koll19b}, who found that for 10 bar atmospheres the day- to nightside heat transport for moderately irradiated tidally-locked planets is very efficient. In their eq.~(9), they provide an approximation for $f$, parameterized by the longwave optical depth at the surface, the surface pressure and the equilibrium temperature. Using our values, we find $f < 0.3$ for all set-ups in this work.

We further incorporate a new addition to the \texttt{HELIOS} radiative transfer scheme.  We include the radiative effects of a (presumed solid) planetary surface, using a constant surface albedo $A_{\rm surf} = 0.1$. This allows us to properly account for optically thin spectral windows and to treat the surface feedback on the atmospheric temperature structure.  The formalism of the surface treatment is described in Appendix \ref{app:surface}. We consider any interior remnant heat to be negligible and set it to zero.

Our stellar spectra are taken from the \texttt{PHOENIX} online library \citep{husser13} and interpolated for the stellar temperatures in Table~\ref{tab:parameters}, $\log_{10} g_{\rm star} = 5.22^y, 5.06^y, 5.06$, and $[M/{\rm H}]_{\rm star} = 0.04, -0.12, 0$ for TRAPPIST-1b \citep{delrez18}, GJ~1132 \citep{berta15} and LHS 3844 \citep{vanderspek19}, respectively. The values marked with $y$ are our own calculations using Newton's law of gravity. The metallicity of LHS~3844 is unknown and thus set to solar.

\section{Results}
\label{sec:res}

\subsection{Temperature Profiles and Spectra}
\label{sec:planet_grid}

\begin{figure*}
\begin{center}
\begin{minipage}{\textwidth}
\includegraphics[width=\textwidth]{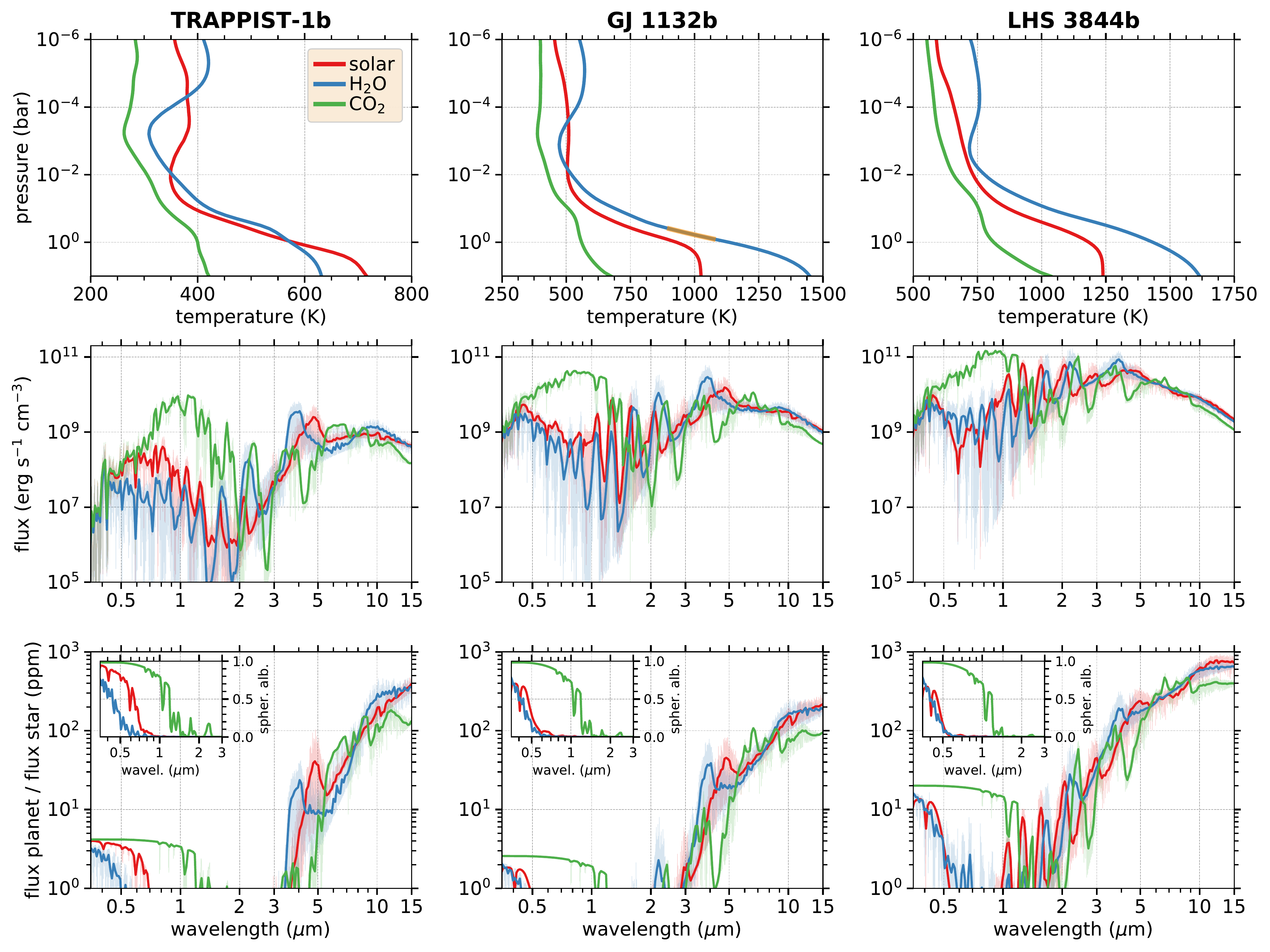}
\end{minipage}
\vspace{-0.3cm}
\caption{{\bf Top row}: dayside temperature profiles for TRAPPIST-1, GJ 1132b, and LHS 3844b for the three atmospheric compositions, 100\% H$_2$O, 100\% CO$_2$, and equilibrium chemistry with solar abundances. A convective zone (orange) appears in the H$_2$O model for GJ 1132b. {\bf Middle row}: planetary emission spectra for the three planets explored. {\bf Bottom row}: secondary eclipse spectra for the three planets explored. {\bf Inlaid:} spherical albedo versus wavelength. Longward of 3 $\mu$m the spherical albedo is essentially zero in all cases. The color scheme is the same in all panels.}
\label{fig:planet_grid}
\end{center}
\end{figure*}

\begin{figure*}
\begin{center}
\begin{minipage}{\textwidth}
\includegraphics[width=\textwidth]{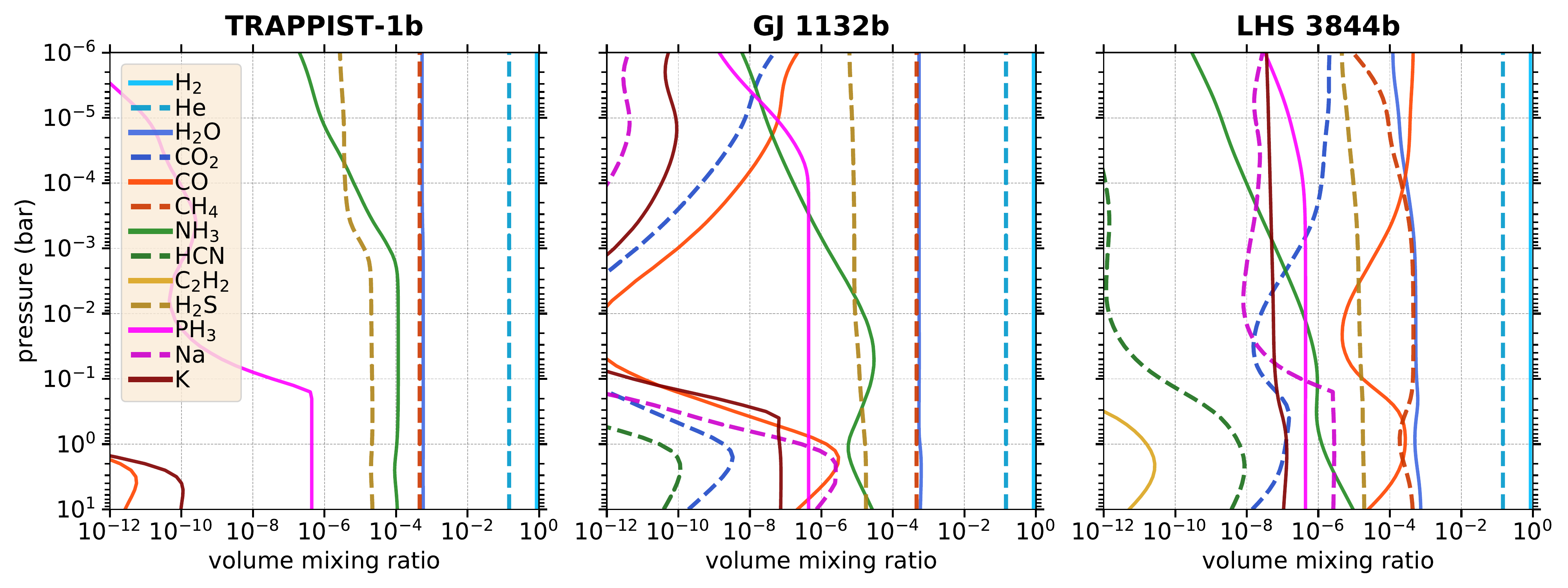}
\end{minipage}
\vspace{-0.3cm}
\caption{Vertical volume mixing ratios for the atmospheric species included in the solar elemental abundance models for the three planets explored. The color scheme is the same in all panels.}
\label{fig:abundances}
\end{center}
\end{figure*}

The top, middle and bottom panels of Fig.~\ref{fig:planet_grid} show the temperature profiles in radiative-convective equilibrium, the emission spectra, and the secondary eclipse spectra for the three test planets and the three atmospheric models, respectively. Inlaid in the bottom panels, the spherical albedo indicates what fraction of the total planetary spectrum is due to the reflected stellar flux. We find that the model atmospheres feature the following characteristics.
\begin{itemize}\itemsep1.5pt
\item{Six of the nine calculated temperature profiles exhibit clear inversions in the upper atmospheric regions, $P < 10$ mbar. They are predominantly found in the TRAPPIST-1b models and to a lesser extent in the GJ 1132b and LHS 3844b models, in this order. In terms of atmospheric composition, inversions appear to be favored in the H$_2$O models, followed by the CO$_2$ and the solar models.}
\item{Only one of the nine simulated atmospheres possesses a convective zone, even though strong greenhouse gases such as H$_2$O and CO$_2$ are present.}
\item{The CO$_2$ atmospheres are significantly cooler than their counterparts of H$_2$O or solar composition. The relative strengths of Rayleigh scattering and surface reflection (due to the non-zero surface albedo), to atmospheric absorption leads to a larger fraction of stellar light being reflected and a smaller fraction being absorbed and converted to heat. Consequently, the CO$_2$ atmospheres also show a stronger reflectance spectrum than the other two cases. In fact, averaging over the spectra of the three planets, we obtain $A_{\rm B}$ = 0.004, 0.29, 0.01 for the Bond albedos of the H$_2$O, CO$_2$ and the solar models, respectively.}
\end{itemize}
Above all, we find prevalent, temperature inversions in our models, which have not been reported in previous parameterized models of similar super-Earth atmospheres \citep[e.g.][]{miller-ricci09, morley17}, but are consistent with the findings of the habitability studies on terrestrial planets of \citet{lincowski18, meadows18}.

Note that the utilized \texttt{PHOENIX} stellar models do not consider the heterogeneous nature of M star photospheres. In particular, not including bright spots likely leads to overpredicting the secondary eclipse signal at shorter wavelengths. \citep{gully-santiago17, morris18}.

In Fig.~\ref{fig:abundances}, we show the vertical mixing ratios of the atmospheric species in the solar abundance models. For TRAPPIST-1b, H$_2$O, CH$_4$, NH$_3$ and H$_2$S, followed by PH$_3$, dominate as absorbers, with the other species playing minor roles. The hotter planets GJ 1132b and LHS 3844b, exhibit a larger pool of major absorbers, with CO$_2$, CO, Na and K also contributing non-negligibly. For all planets, the background gases H$_2$ and He play an active role for the radiative transfer via Rayleigh scattering and CIA absorption.

In the next section, we explore the parameter regime where temperature inversions occur and seek a physical explanation.

\subsection{Emergence of a Temperature Inversion}
\label{sec:inv}

\begin{figure*}
\begin{center}
\begin{minipage}{0.49\textwidth}
\includegraphics[width=\textwidth]{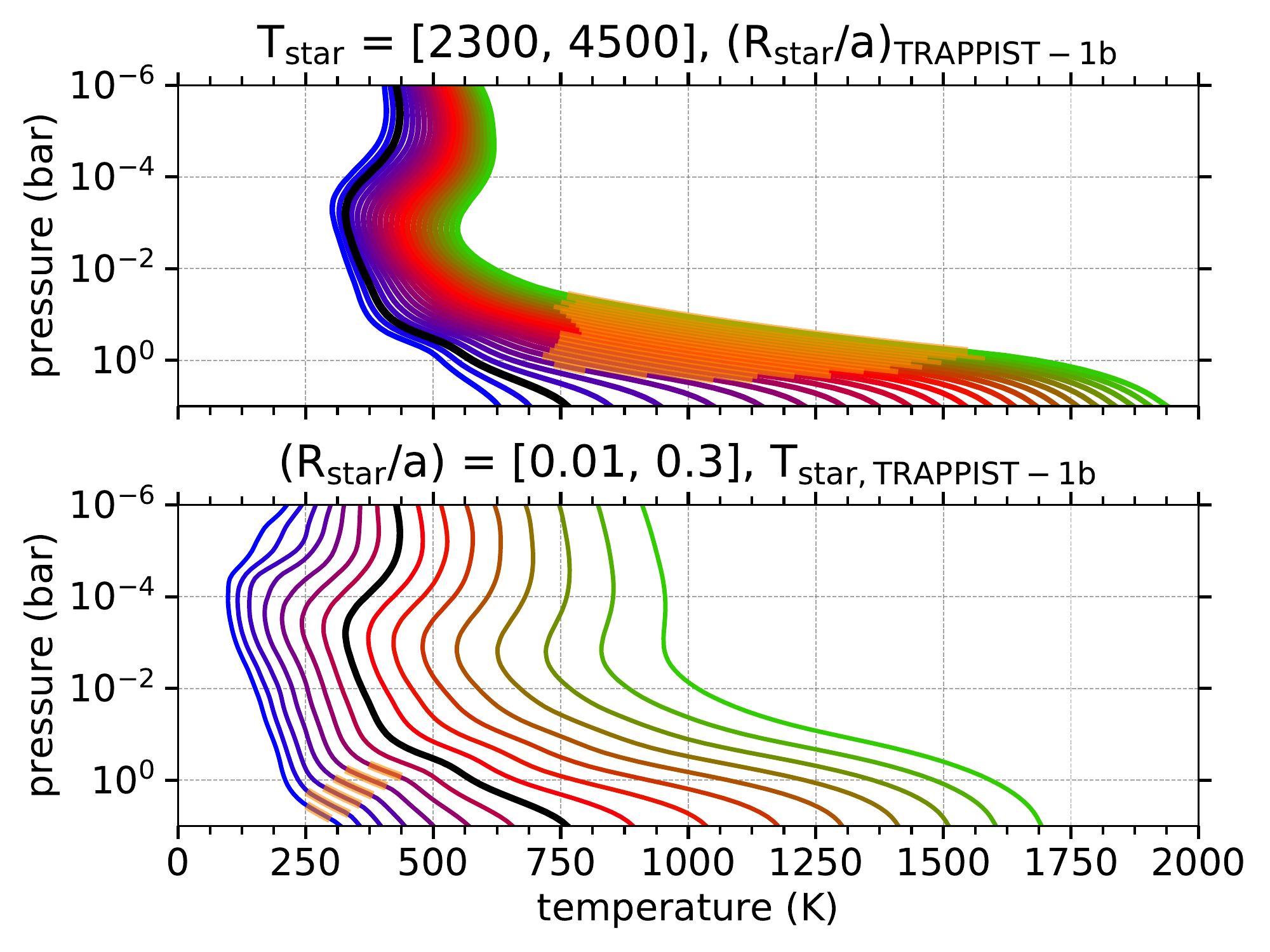}
\end{minipage}
\hfill
\begin{minipage}{0.49\textwidth}
\includegraphics[width=\textwidth]{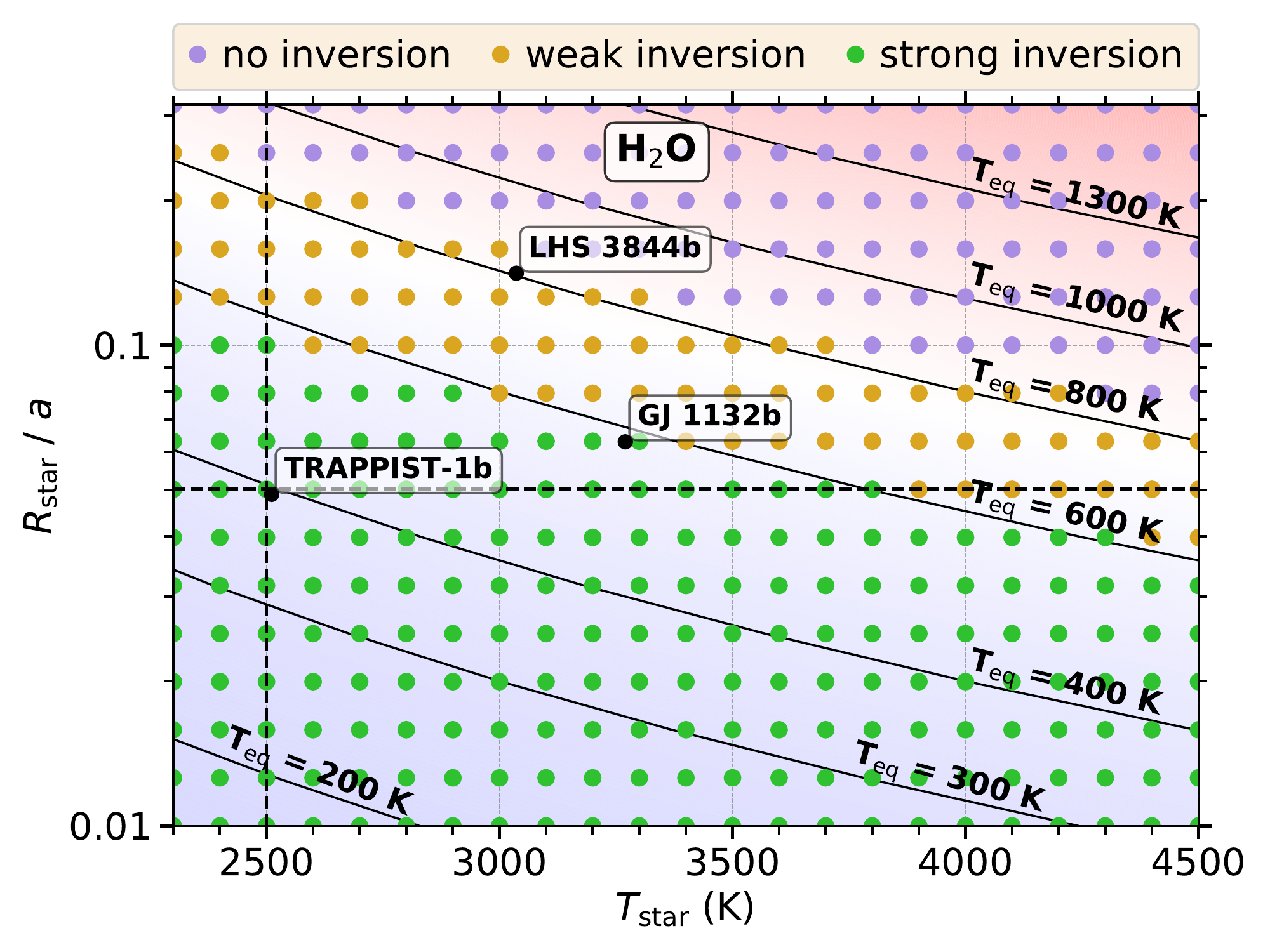}
\end{minipage}
\begin{minipage}{0.49\textwidth}
\includegraphics[width=\textwidth]{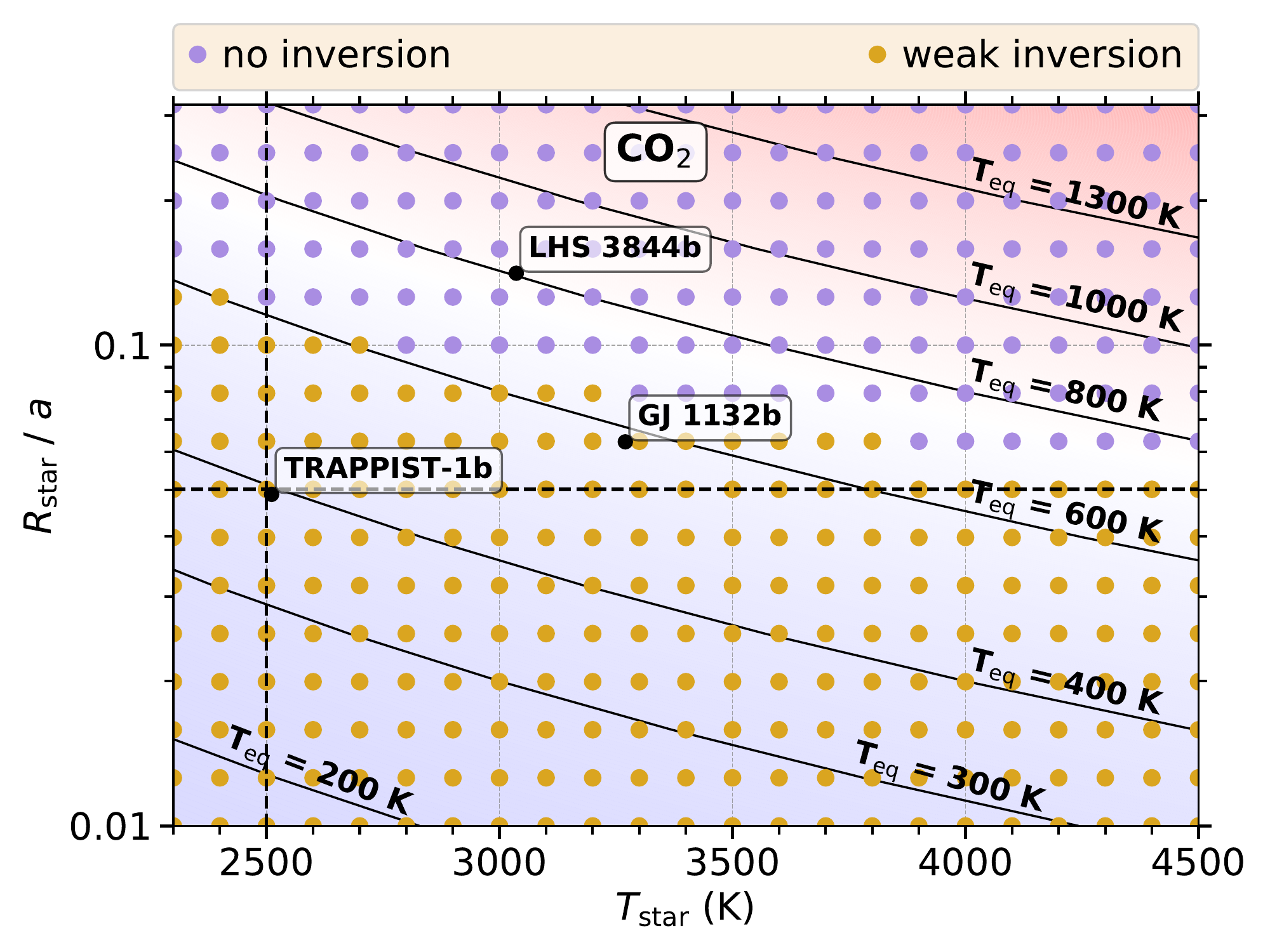}
\end{minipage}
\hfill
\begin{minipage}{0.49\textwidth}
\includegraphics[width=\textwidth]{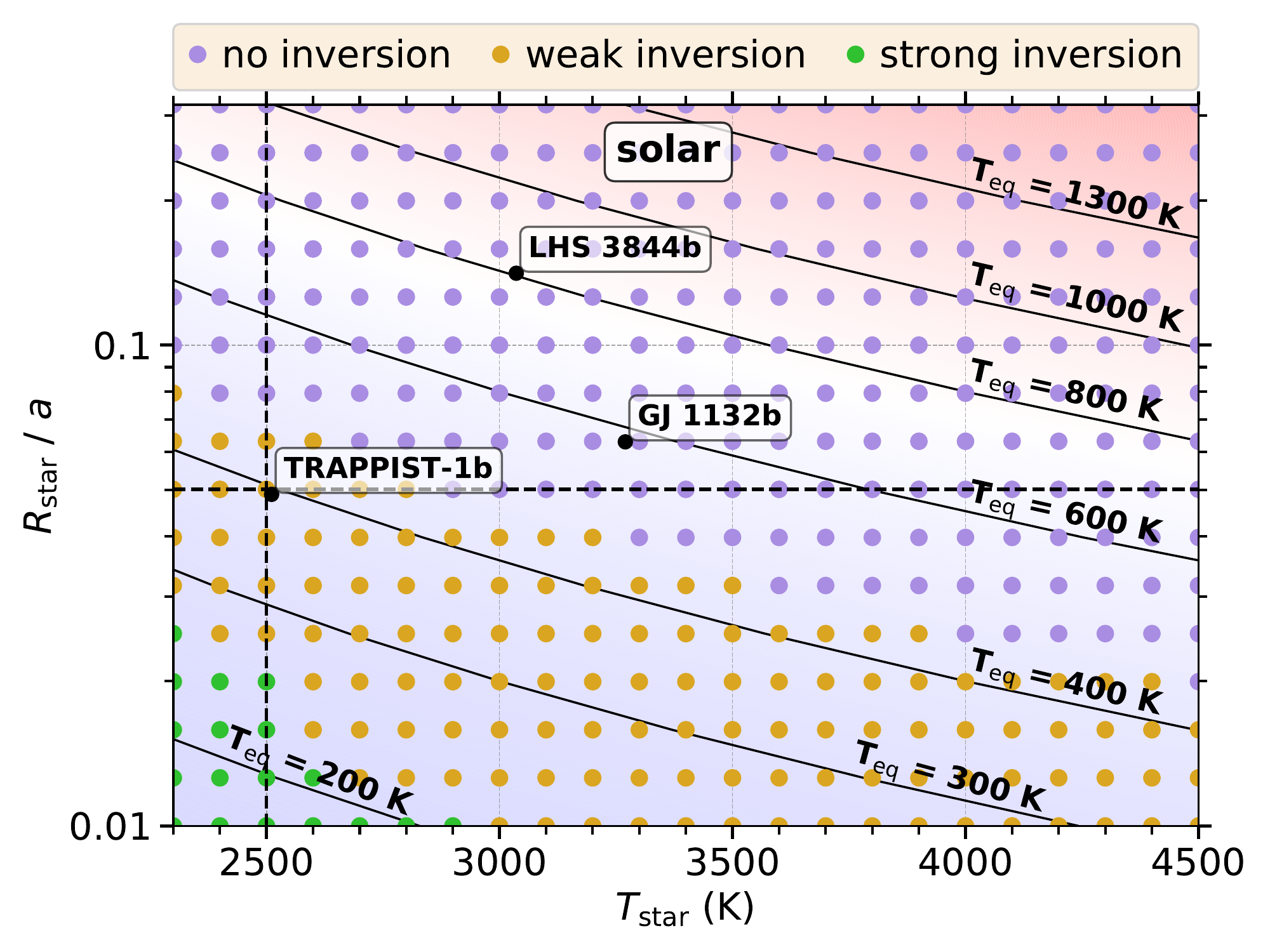}
\end{minipage}
\vspace{-0.3cm}
\caption{{\bf: Top left:} grid of dayside temperature profiles for a TRAPPIST-1b-like planet with changing stellar temperature (top) and stellar radius over orbital distance ratio (bottom). The atmospheric composition is 100\% H$_2$O. Convective regions are displayed in thick orange. The model with parameters closest to TRAPPIST-1b is marked in black. {\bf Other panels:} emergence of a temperature inversion versus stellar temperature and stellar radius over orbital distance for the three explored atmospheric compositions: 100\% H$_2$O (top right), 100\% CO$_2$ (bottom left), solar abundances (bottom right). A temperature inversion is marked strong (weak) if the maximum temperature is more (less) than 20 \% higher than the minimum temperature in the inversion layers. Overlayed are the corresponding equilibrium temperatures for an assumed Bond albedo of 0.1. Also, the locations of the three test planets in terms of those parameters are marked. The dashed lines in the top right panel represent the model slices shown on the top left.}
\label{fig:inv_contour}
\end{center}
\end{figure*}

Neglecting any interior heat, the atmospheric radiation is effectively determined by one external factor, the incoming stellar flux. At the top-of-atmosphere (TOA) the downward flux is given by the Stefan-Boltzmann law as
\begin{equation}
\mathcal{F}_{\downarrow, \rm TOA} = \left(\frac{R_{\rm star}}{a}\right)^2 \sigma_{\rm SB} T_{\rm star}^4, 
\end{equation}
where $R_{\rm star}$ is the stellar radius, $a$ is the orbital distance, $\sigma_{\rm SB}$ is the Stefan-Boltzmann constant and $T_{\rm star}$ is the stellar temperature. Assuming a blackbody spectrum for the star, the quantities $(R_{\rm star}/a)$ and $T_{\rm star}$ uniquely determine the amount and the wavelength distribution of the external irradiation. The atmospheric heat content for a planet (in global energy balance) is usually parameterized by the equilibrium temperature
\begin{equation}
T_{\rm eq} = \left(\frac{1 - A_{\rm B}}{4}\right)^{0.25} \left(\frac{R_{\rm star}}{a}\right)^{0.5} T_{\rm star} ,   
\end{equation}
where $A_{\rm B}$ is the planet's Bond albedo. From an atmospheric perspective, the only ``free'' parameters are $(R_{\rm star}/a)$ and $T_{\rm star}$ as before. The albedo $A_{\rm B}$ is an intrinsic property of the atmosphere and surface and is a consequence of the intricate interplay between matter and radiation. Thus, $A_{\rm B}$ cannot be chosen independently of the atmospheric model. We hence conduct a two-dimensional parameter space exploration in $(R_{\rm star}/a)$ and $T_{\rm star}$ using the three atmospheric models ($100\%$ H$_2$O, $100\%$ CO$_2$ and solar composition) and the surface gravity $\log_{10} g = 3.05$. As before, we assume $A_{\rm surf} = 0.1$. Furthermore, we set up the star as a blackbody (for the models shown in Fig.~\ref{fig:inv_contour} only), in order to exclude the intricate wavelength-dependence of a realistic star. We have found the latter choice to have a minor effect on the qualitative shape of the atmospheric temperature profile (not shown).

Fig.~\ref{fig:inv_contour} shows which combinations of values in our calculations lead to atmospheric temperature inversions, where $d T / d P < 0$. Consistent with our previous findings, the temperature inversions are more ubiquitous in a wider parameter regime for H$_2$O models than for CO$_2$ models and solar composition models. In addition, H$_2$O atmospheres often feature stronger, more pronounced inversions than the other two cases. The upper left panels of Fig.~\ref{fig:inv_contour} show the temperature profiles of slicing through the H$_2$O model grid, once with changing $T_{\rm star}$ and once with changing $(R_{\rm star}/a)$, intersecting at the model closest to TRAPPIST-1b. Interestingly, we find that reducing $(R_{\rm star}/a)$ has a larger impact on the emergence of temperature inversions than increasing $T_{\rm star}$. This leads to the conclusion that the prevailing atmospheric temperatures are more important than the stellar peak emission in this regard. In general, the boundary between models with and without inversions roughly follows the equilibrium temperature contours, supporting the result that the atmospheric heat content is the major factor at hand, with the stellar temperature playing a secondary role. Note that the equilibrium temperatures are approximated assuming $A_{\rm B} = 0.1$. In reality, the Bond albedo varies between the individual models. For optically thin atmospheres, we expect $A_{\rm B} \gtrsim A_{\rm surf} = 0.1$, but $A_{\rm B} < A_{\rm surf}$ with increasing optical thickness. In the following, we offer an explanation for our results.

\begin{figure*}
\begin{center}
\begin{minipage}{0.49\textwidth}
\includegraphics[width=\textwidth]{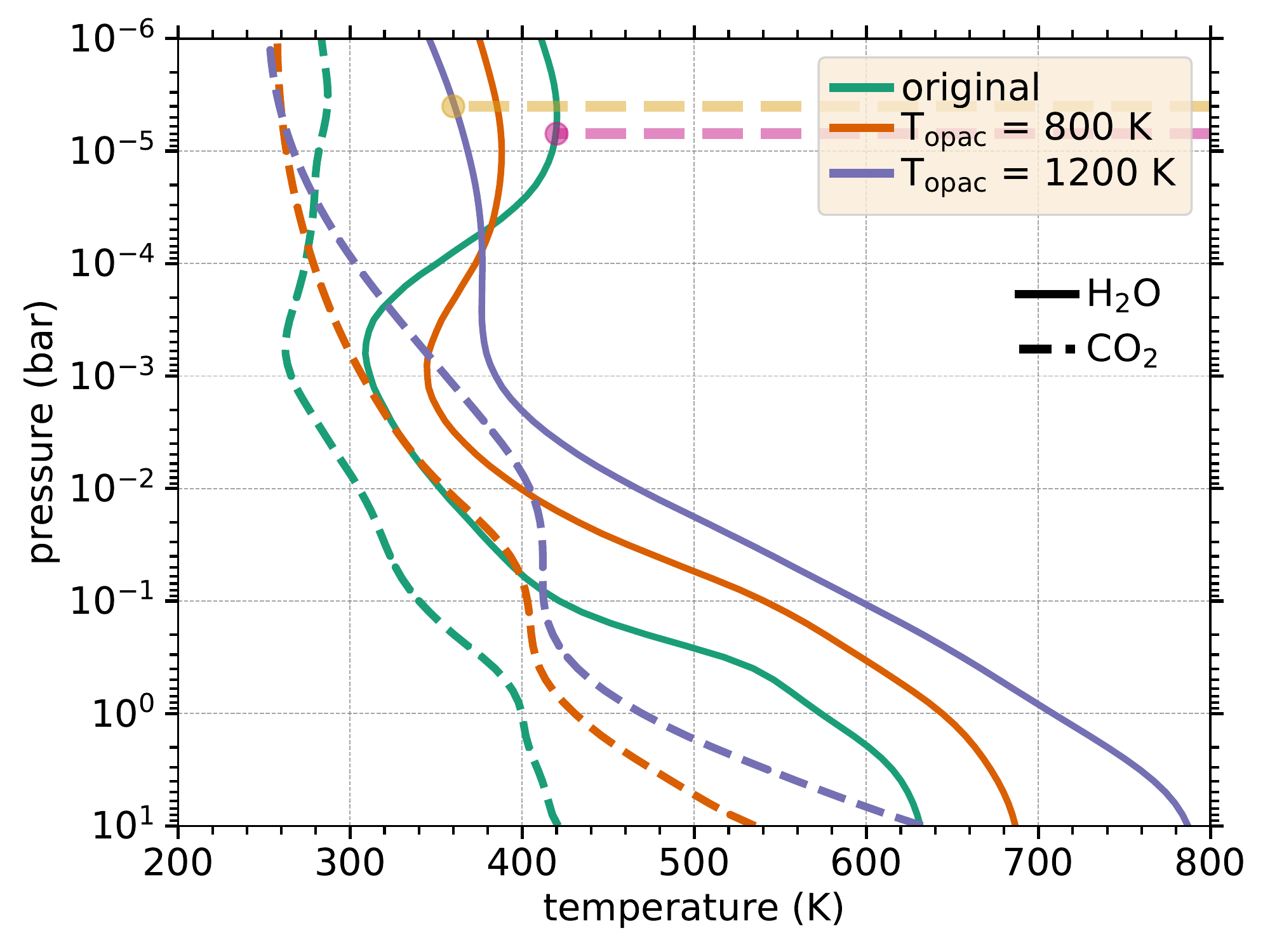}
\end{minipage}
\hfill
\begin{minipage}{0.49\textwidth}
\includegraphics[width=\textwidth]{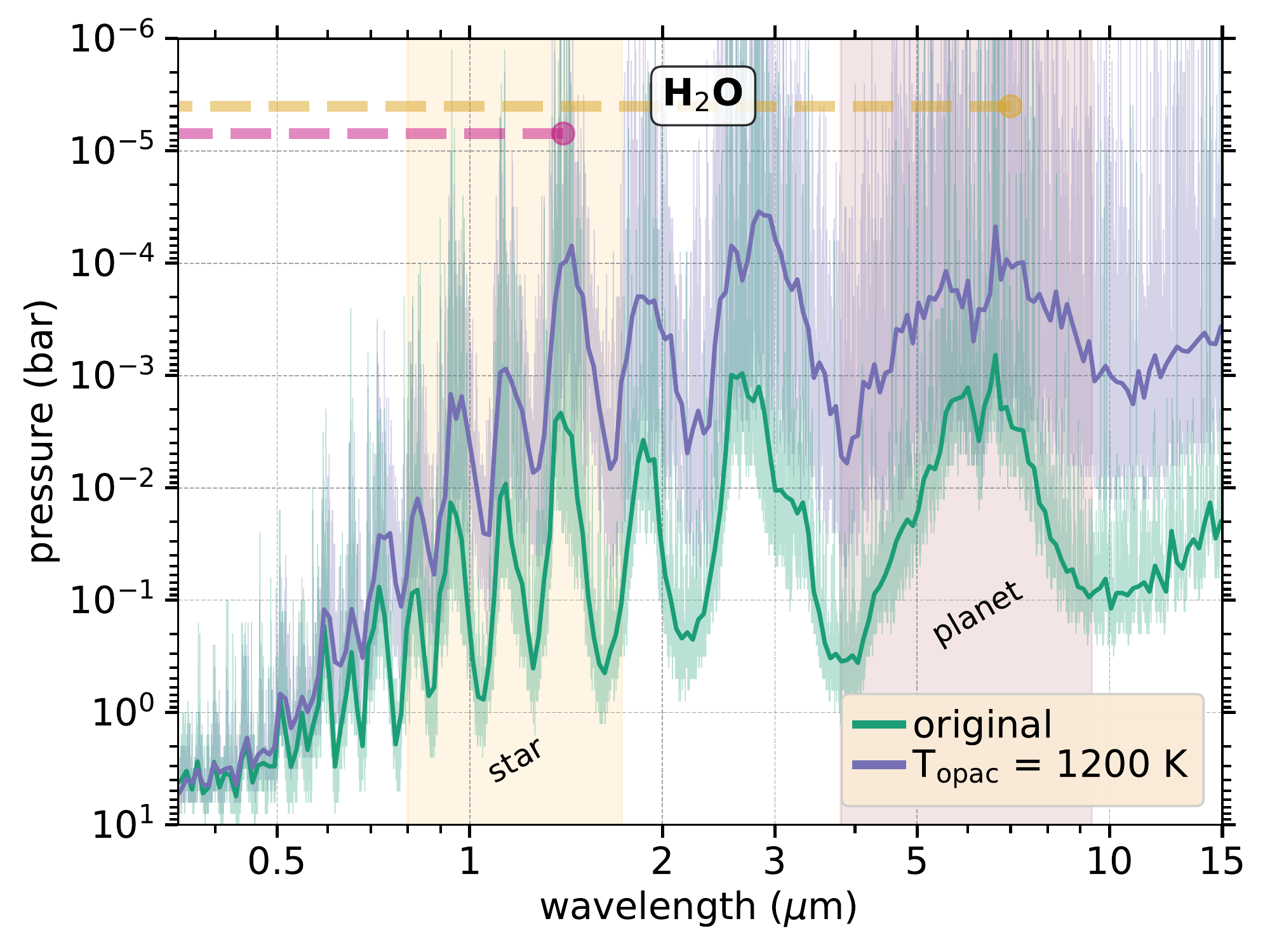}
\end{minipage}
\begin{minipage}{0.49\textwidth}
\includegraphics[width=\textwidth]{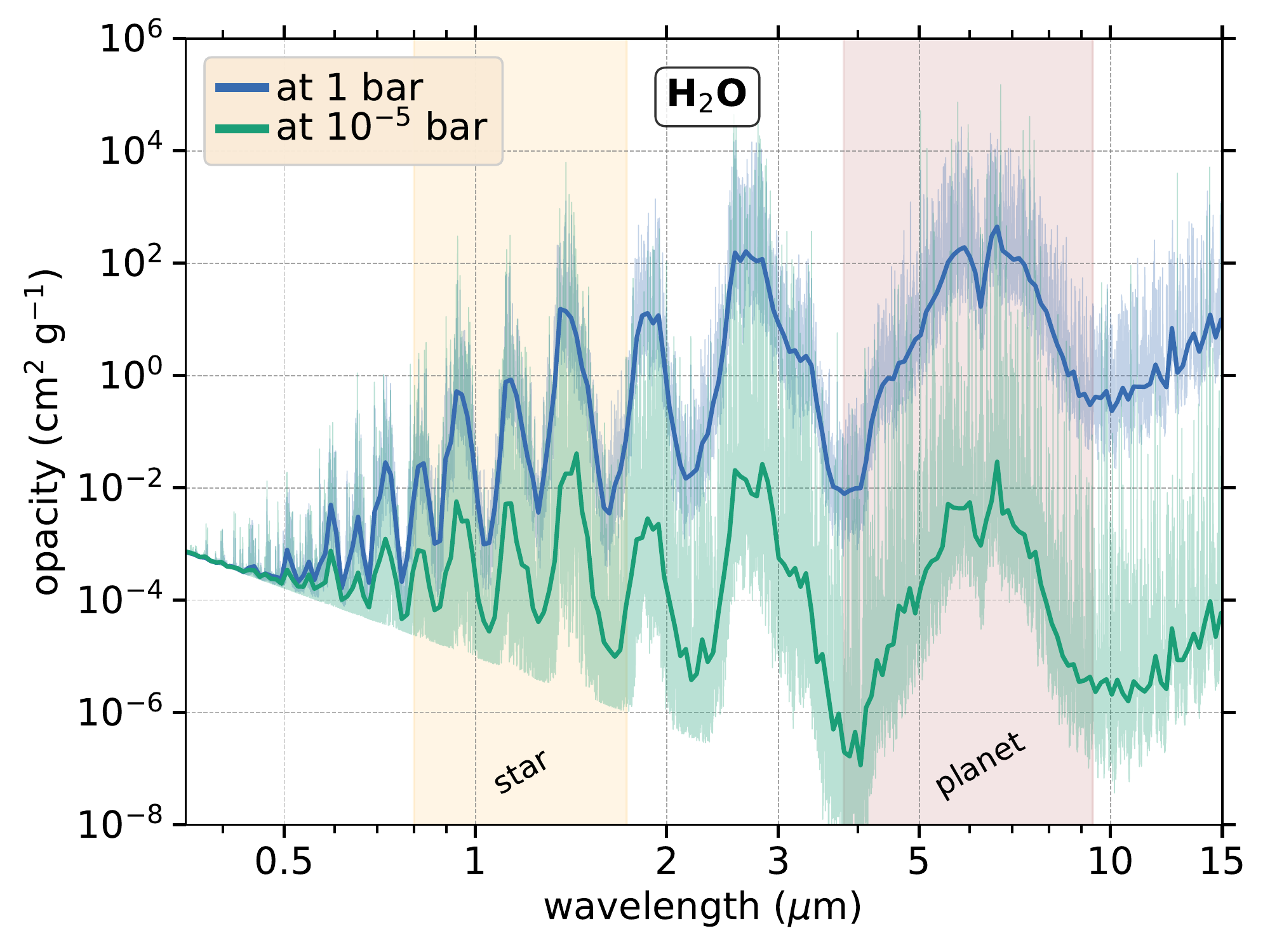}
\end{minipage}
\hfill
\begin{minipage}{0.49\textwidth}
\includegraphics[width=\textwidth]{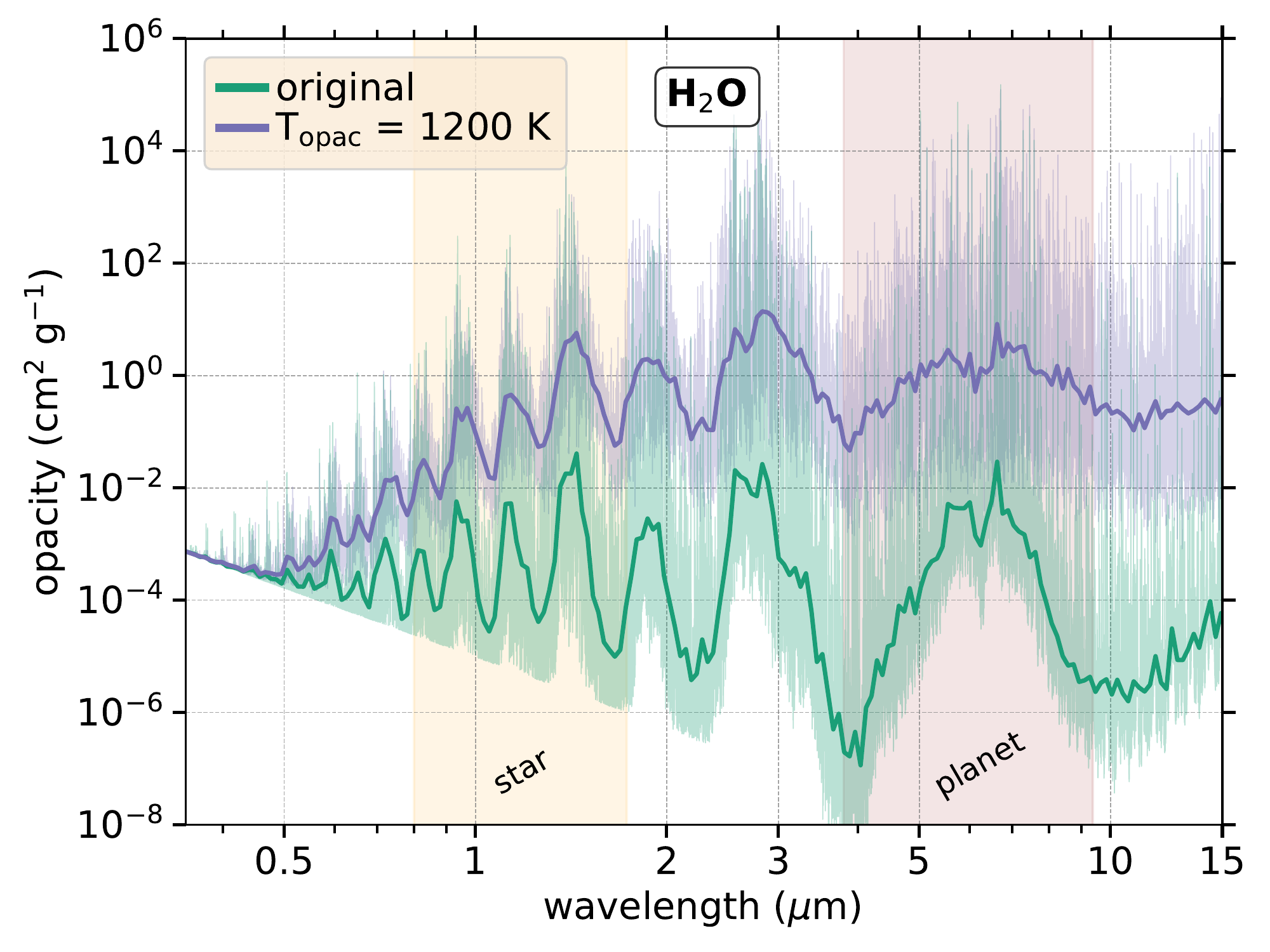}
\end{minipage}
\vspace{-0.3cm}
\caption{{\bf Top left:} temperature profiles for TRAPPIST-1b for the pure H$_2$O (solid) and pure CO$_2$ (dashed) atmospheres. The original profiles (from Fig. \ref{fig:planet_grid}) are compared to two test simulations, which use opacities calculated at the fixed temperatures of 800 K and 1200 K. {\bf Top right:} location of the maximum of the transmission weighting function (equal $\approx$ photosphere) for the original and $T_{\rm opac} = 1200$ K simulations. The yellow and magenta lines enhance visibility of the connection between the shape of the transmission weighting function and the temperature profiles. The pure H$_2$O case is shown in the top right and bottom panels. {\bf Bottom left:} opacities at the atmospheric pressures of 1 bar and 10$^{-5}$ bar, shown for the original model. {\bf Bottom right:} opacities for the original and $T_{\rm opac} = 1200$ K models. In the top right, bottom left and bottom right panels, the quantities are shown both in $R=3000$ (shaded) and $R=50$ (solid line) resolution. The shaded regions in the background show the approximate wavelengths encapsulating $\sim$ half of the stellar and planetary radiation.}
\label{fig:inv_expl}
\end{center}
\end{figure*}

In contrast to hotter stars, e.g., the Sun, the bulk emission of an M star happens at longer, NIR wavelengths. For instance, TRAPPIST-1 with a temperature of 2511 K, has its peak emission at 1.2 $\mu$m (Wien's law). There, H$_2$O and CO$_2$ have sufficiently strong absorption bands to absorb a part of the incoming stellar flux high up in the atmosphere. This has a twofold effect. First, a temperature inversion, may be created, even in the absence of gases efficiently absorbing in the optical. Second, due to the increased attenuation of stellar flux, only a smaller fraction of it reaches the lower near-surface layers, muting the accumulation of heat and subsequent convective instability. As result, the lower atmosphere is more isothermal and convective zones are rarer and, if present, less pronounced for planets around M dwarfs than around hotter stars. Note that for thinner atmospheres than explored here, near-surface convection may still arise. In this case, the larger amount of stellar flux deposited at the surface and the corresponding increase in temperature create a larger lapse rate between the surface and the immediate atmospheric layers above.

In Fig.~\ref{fig:inv_expl}, top left, we first focus on the solid, green line representing the temperature profile in the H$_2$O model for TRAPPIST-1b. On the top right, the pressure level of the wavelength-varying photosphere is shown. For an atmospheric model with $n$ vertical layers, the location of the photosphere is calculated using the transmission weighting function $\Psi$, which at the $i$-th layer is given by
\begin{equation}
\Psi_i = (1 - \mathcal{T}_i) \prod_{\substack{j=i+1 \\ i \neq n-1}}^{n-1} \mathcal{T}_j ,
\end{equation}
with $\mathcal{T}$ being the transmission function over the width of the respective layer. To first order, the photosphere coincides with the atmospheric layer where $\Psi$ is maximal. The transmission weighting function is closely related to the commonly used contribution function \citep[e.g.,][]{irwin09}, but without taking the atmospheric thermal emission into account. At the wavelengths of the stellar irradiation, around 1 - 2 $\mu$m the strong H$_2$O line cores makes the photosphere extend nearly to the TOA and thus the atmosphere is able to absorb the stellar energy there (connecting magenta line). Note that the transmission weighting function, down-sampled at resolution R=50, misses the highest peaks, originating from strong H$_2$O lines. In this low-resolution case, the photosphere is located significantly below the found inversion layers. Thus, models at such a resolution might substantially underpredict the occurrence of high altitude inversions.

Fig.~\ref{fig:inv_expl} also shows why the inversion disappears when increasing the atmospheric temperatures. We find this behavior to be predominantly an opacity effect, namely a consequence of the non-gray band structure of the opacities. In the top left panel, we show test models, for which the opacities are pre-calculated at fixed temperatures, 800 and 1200 K, independent of the actual atmospheric conditions. With higher temperature more spectral lines become active, filling up the spectral bands and increasing the overall opacity, particularly in the infrared (see bottom right). This amplification of absorption and emission shifts the photosphere higher up in the infrared, allowing the atmosphere to sufficiently cool and thus preventing a temperature inversion (connecting yellow line). We have tested and found this behavior to be true for both H$_2$O and CO$_2$ models.

We would like to further highlight that the overall slope of the opacity function depends on the pressure (see Fig.~\ref{fig:inv_expl}, bottom left). While in the lower atmosphere the H$_2$O opacity has an increasing slope with wavelength, higher up this trend vanishes. Hence the propensity for H$_2$O to absorb at the peak of the stellar blackbody increases at higher altitudes, leading to temperature inversions in atmospheres that are optically thick at those locations.

Our observed trend on the emerge of temperature inversions in the atmospheres of rocky planets contrasts the appearance of temperature inversions in hot Jupiters. In the latter case, only very hot conditions are expected to lead to temperature inversions, through the shortwave absorption bands of metal oxides/hydrides or the lack of water emission due to its dissociation \citep[e.g.][]{fortney08, arcangeli18, malik19}.

\section{Discussion}
\label{sec:disc}

\subsection{Observability}
\label{sec:detect}

The {\it James Webb Space Telescope} ({\it JWST}) will be the first observation facility to comprehensively record the spectral appearance of rocky exoplanets. In the following, we first discuss whether the different atmospheric models for the three test planets can be distinguished with {\it JWST} based on their emission spectra. Second, we estimate if the temperature inversions found in this study may be observable either with {\it JWST} or ground-based spectroscopy.

\subsubsection{Distinguishing Atmospheric Compositions}
\label{sec:detect_chem}

\begin{figure*}
\begin{center}
\begin{minipage}{0.49\textwidth}
\includegraphics[width=\textwidth]{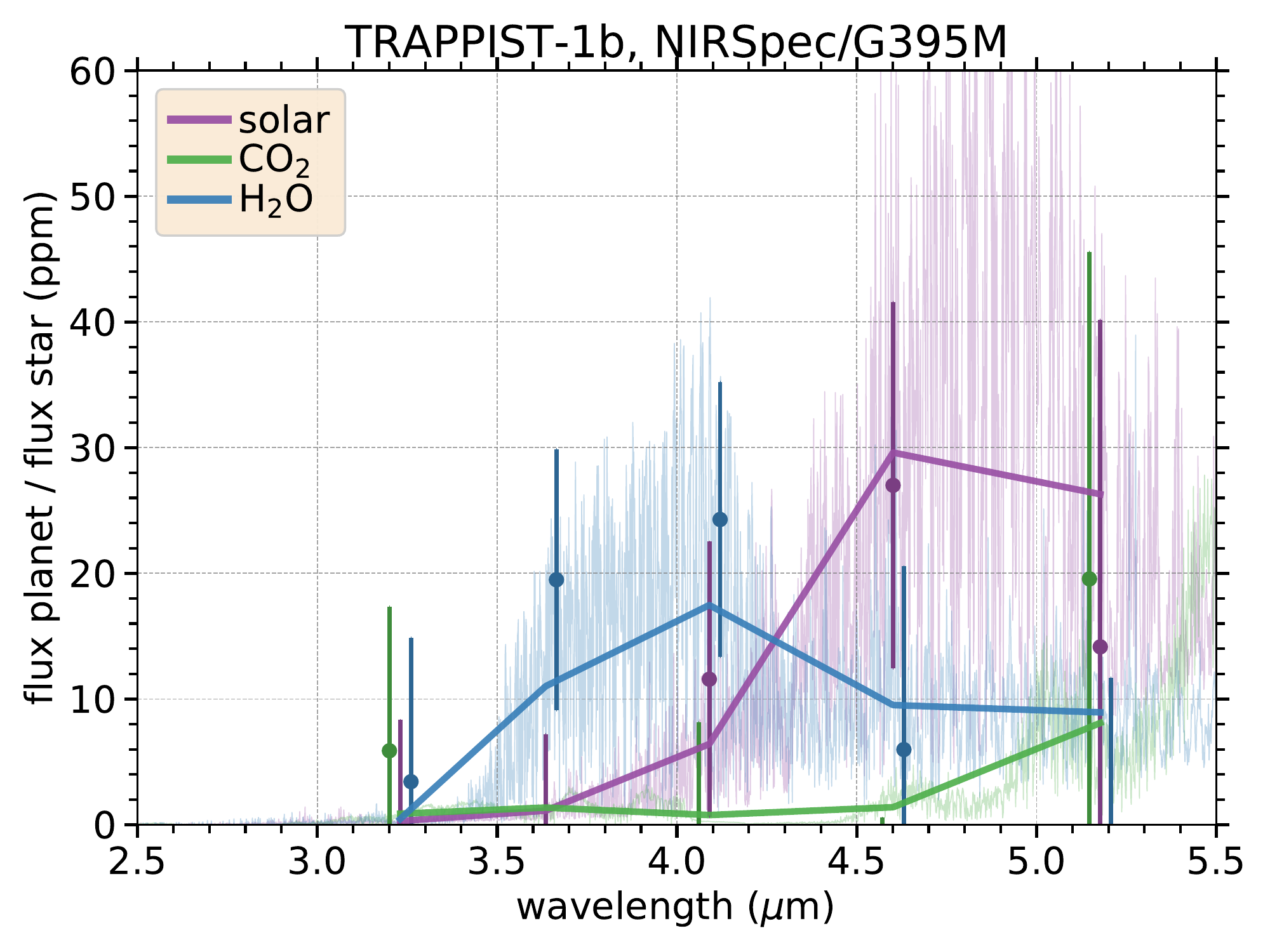}
\end{minipage}
\hfill
\begin{minipage}{0.49\textwidth}
\includegraphics[width=\textwidth]{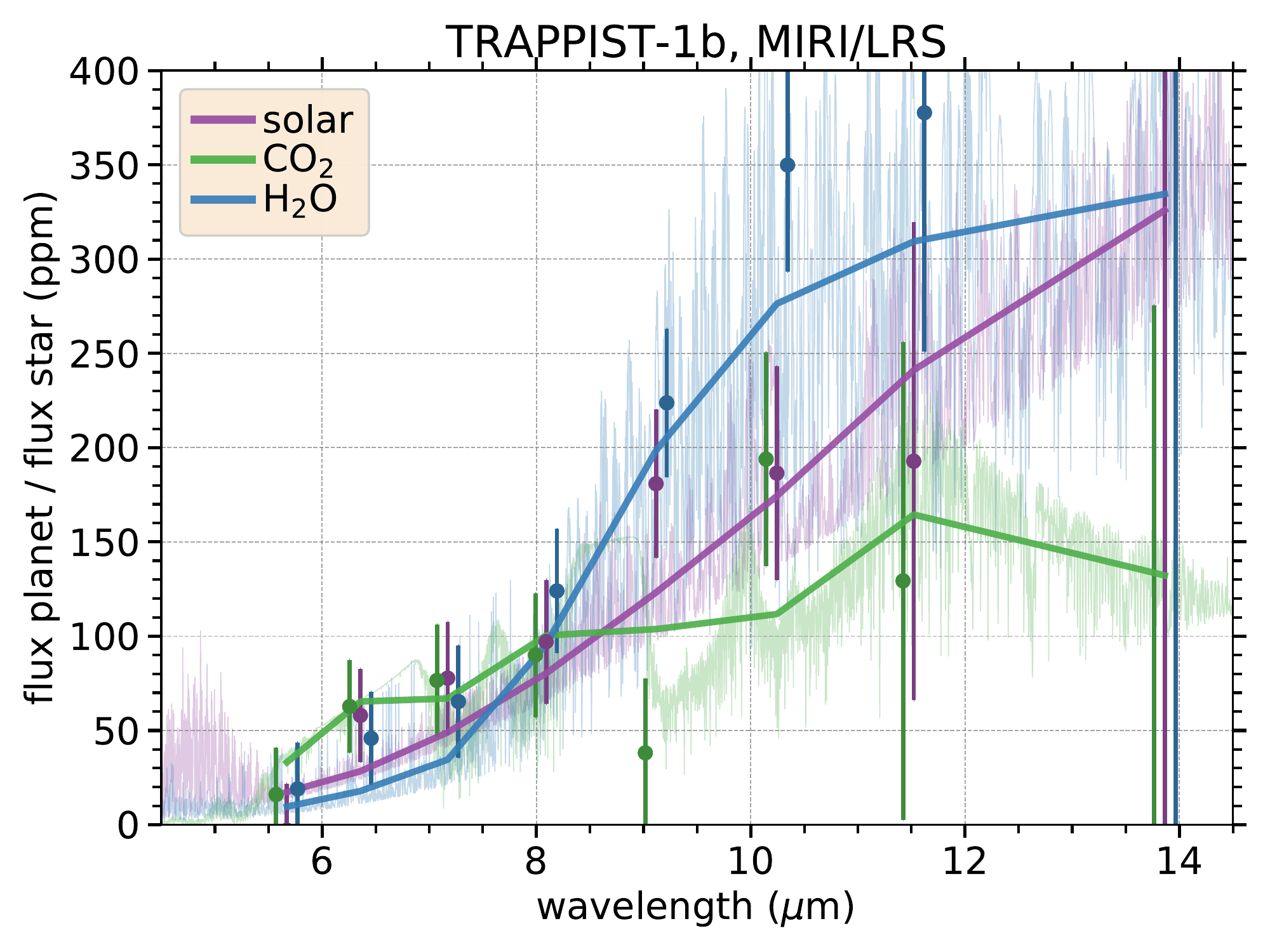}
\end{minipage}
\begin{minipage}{0.49\textwidth}
\includegraphics[width=\textwidth]{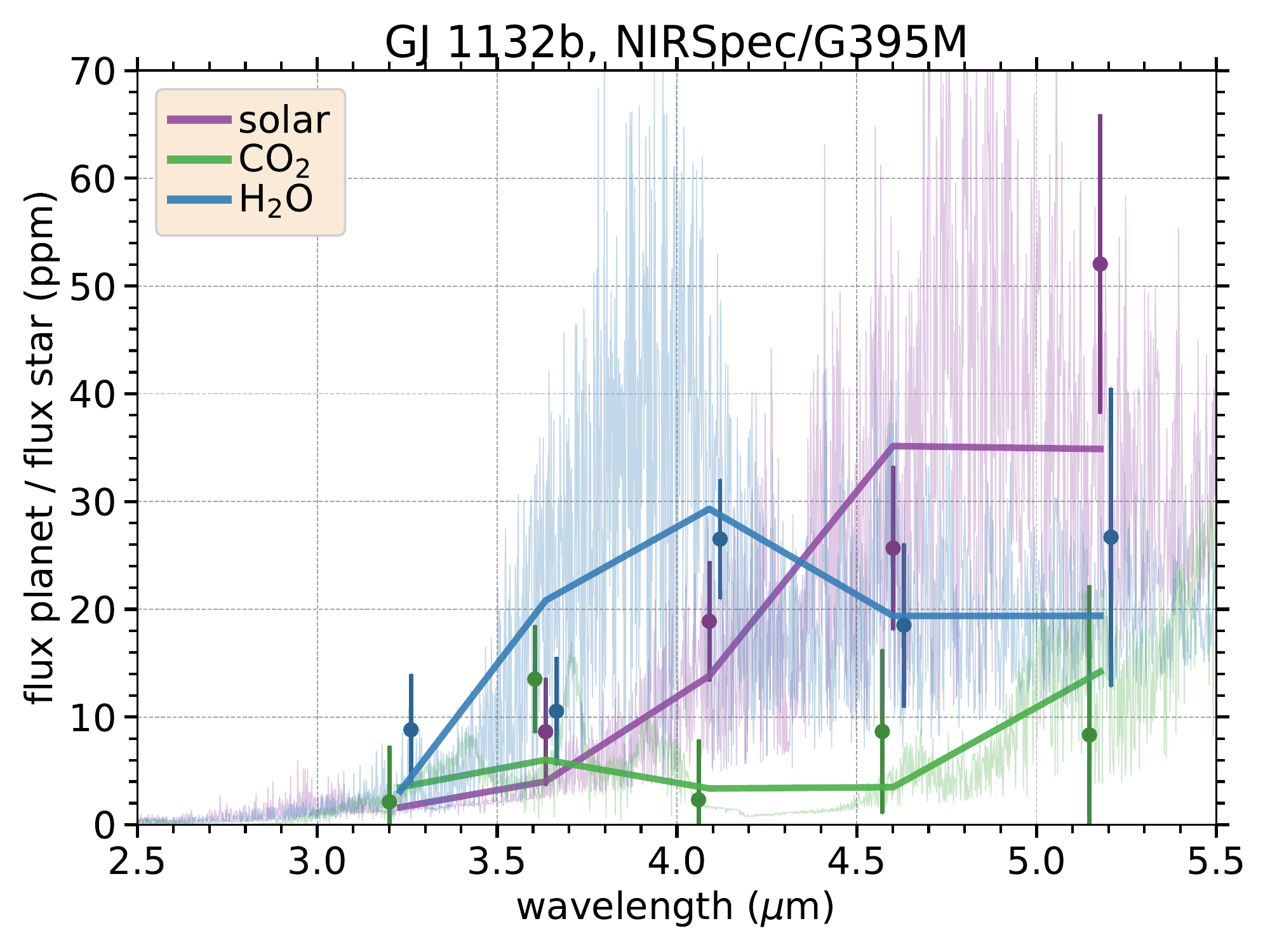}
\end{minipage}
\hfill
\begin{minipage}{0.49\textwidth}
\includegraphics[width=\textwidth]{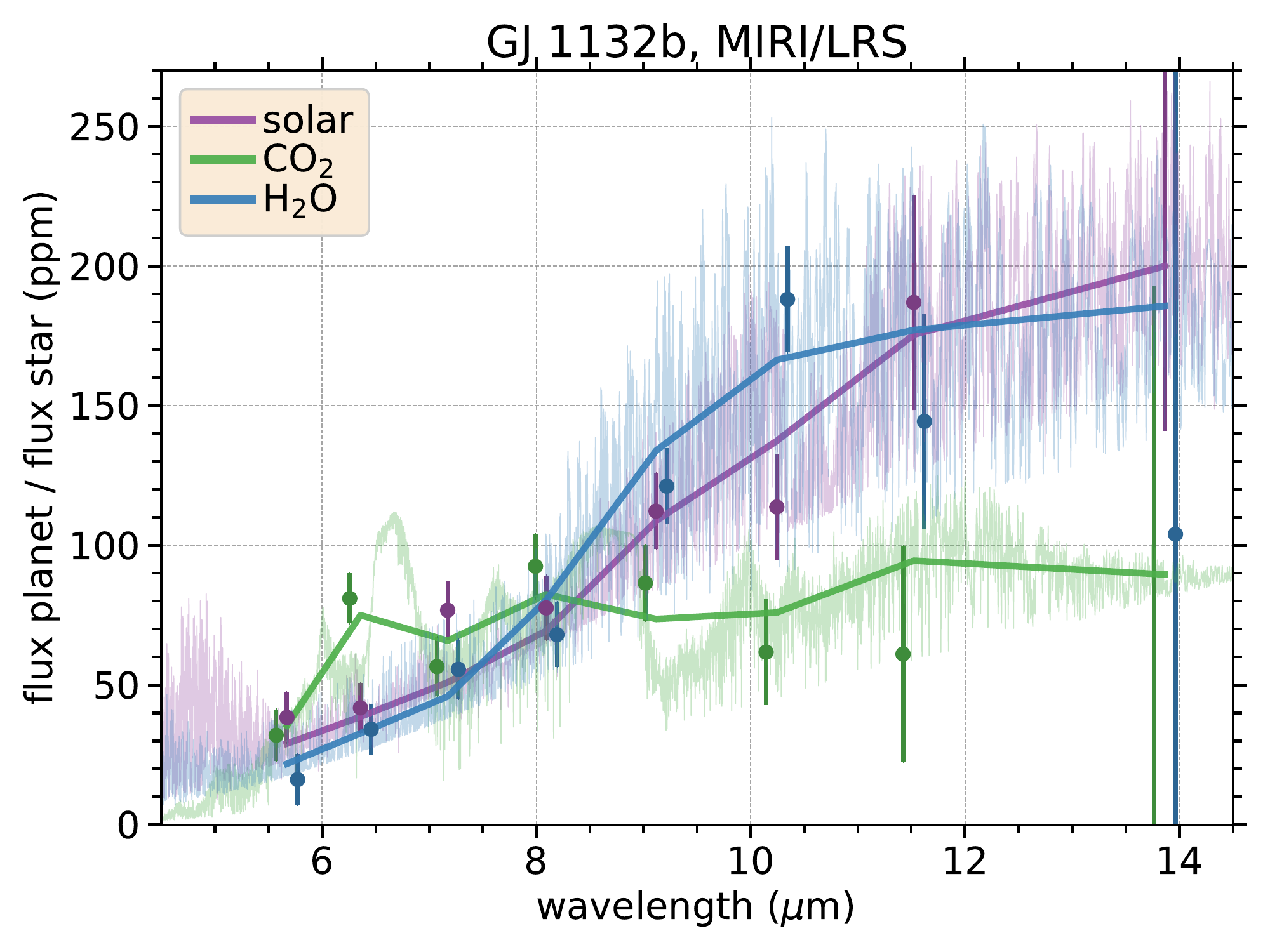}
\end{minipage}
\begin{minipage}{0.49\textwidth}
\includegraphics[width=\textwidth]{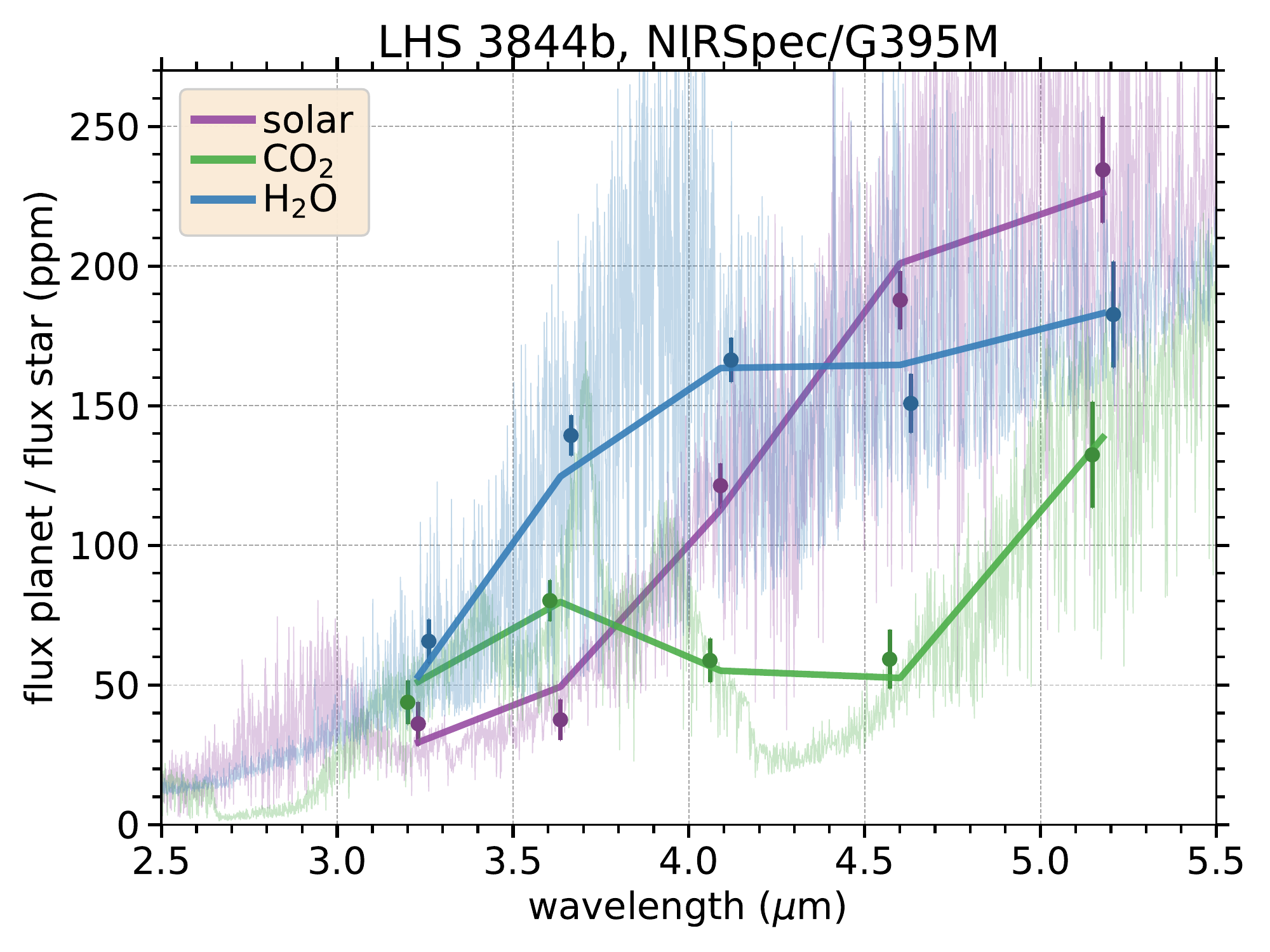}
\end{minipage}
\hfill
\begin{minipage}{0.49\textwidth}
\includegraphics[width=\textwidth]{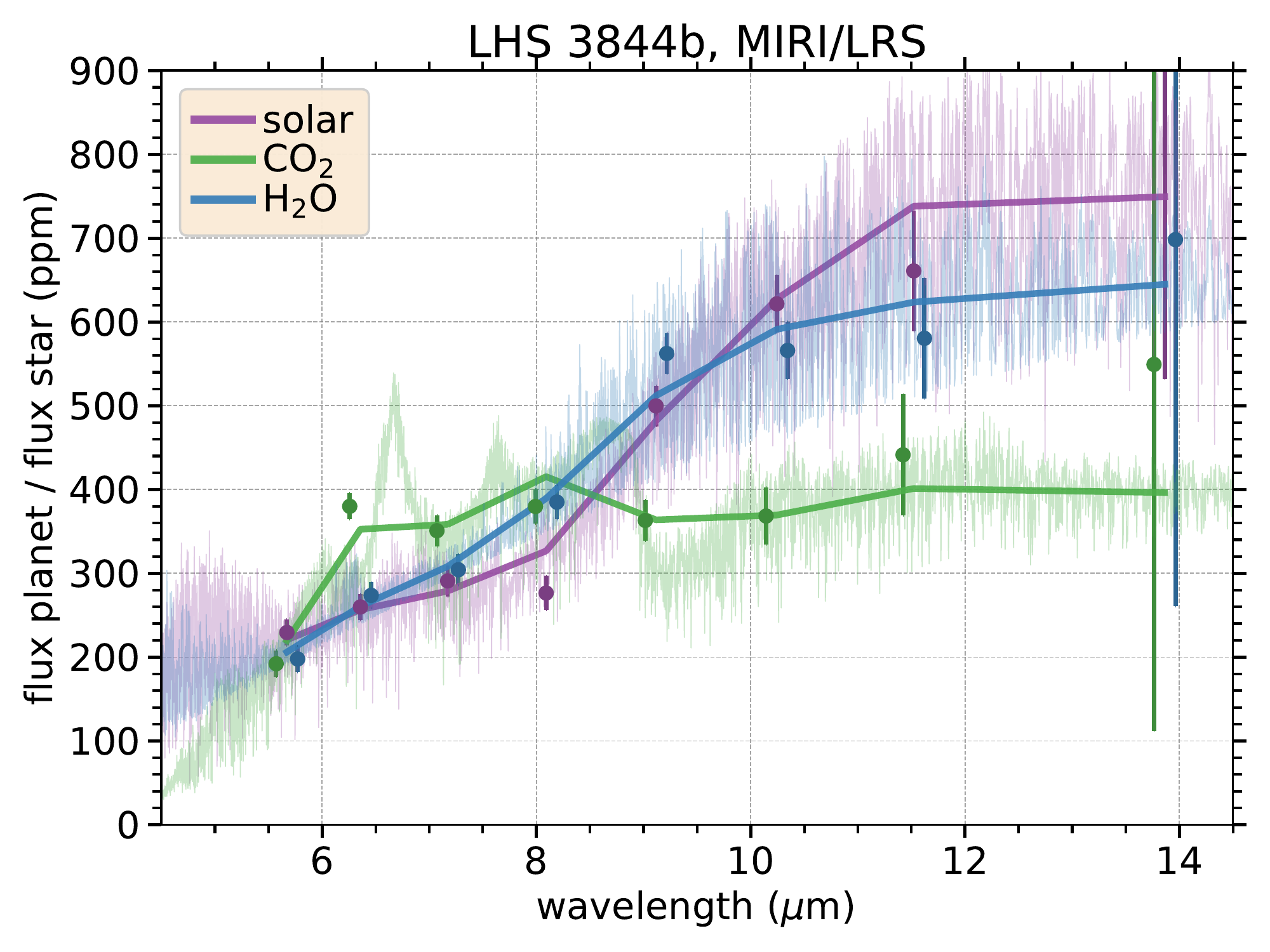}
\end{minipage}
\vspace{-0.3cm}
\caption{Simulated {\it JWST} secondary eclipse spectra for TRAPPIST-1b (top), GJ 1132b (middle), and LHS 3844b (bottom) for the three atmospheric compositions explored. The mock data points with 1$\sigma$ error bars for {\it NIRSpec} (left) and {\it MIRI} (right) are shown for 10-eclipse mock runs and rebinned to the resolution $R=5$. The \texttt{HELIOS} synthetic spectra are shown as both, binned to the resolution of the data (solid lines), and in the original resolution $R=3000$ (shaded). The data points are slightly offset against each other for clarity.}
\label{fig:detect_chem}
\end{center}
\end{figure*}

Using the {\it JWST} instrument noise simulator \texttt{PandExo} \citep{batalha17a} we generate mock secondary eclipse observations of our models. We observe each planet and each atmospheric composition with the NIRSpec G395M mode, and MIRI LRS mode, and determine for various numbers of eclipses the confidence level at which two models are mutually exclusive. We do this by applying the Chi-square test for goodness of fit between mock data and synthetic spectrum of two different models. In terms of instrument parameters, we set the saturation level to $60 \%$ full well and the total observation time to three times the eclipse duration. We obtain the same results for a noise floor of zero or 30 ppm. Testing various binning resolutions, we find that $R=5$ is a reasonable compromise between signal strength and wavelength information content.

Fig.~\ref{fig:detect_chem} shows a snapshot of those results. In the following we assume that models are distinguishable if one model's theoretical values and the other model's simulated data differ by more than 5$\sigma$. 

In the case of LHS 3844b, the different atmospheric models should be distinguishable with only 2 eclipses. Although the overall planet-to-star contrast is larger in the MIRI bandpass, contrasting opacity bands of the different models make NIRSpec the favorable instrument for this planet. The first upward bump in the H$_2$O spectrum at 3.8 $\mu$m is a spectral window in the band structure of H$_2$O.  This window is filled by CH$_4$ absorption in the solar model. The second feature at 4.8 $\mu$m (again seen as an upward bump) is the blue-ward side of a strong H$_2$O absorption band, which is less pronounced in the solar model compared to the H$_2$O model due to the smaller H$_2$O abundance. We expect that the CO$_2$ model should be distinguishable from the other models mainly due to two effects: first, CO$_2$ has a strong absorption band at 4.3 $\mu$m, which is missing in the H$_2$O and solar models, and second, the CO$_2$ model exhibits weaker overall emission due to its generally lower atmospheric temperatures. By enriching the solar model with CO$_2$ we can reproduce the 4.3 absorption feature (not shown), diminishing the ability to distinguish between the solar and CO$_2$ models somewhat. Still, even a solar model enriched with 10$^4\times$ solar value of CO$_2$ can be distinguished from the pure CO$_2$ model with only 4 eclipses for LHS 3844b.

The overall smaller signal of GJ 1132b renders the NIRSpec yield so weak that MIRI takes over as preferred instrument for this planet. With MIRI 5 - 10 eclipses are required to distinguish between the H$_2$O and CO$_2$ models and between the CO$_2$ and solar models. However, since the H$_2$O and solar models exhibit very similar eclipse spectra in the MIRI bandpass, we find NIRSpec to fare better with $\sim 15$ eclipses needed to distinguish between these two scenarios, in contrast to $\gtrsim 30$ eclipses using MIRI.

For TRAPPIST-1b, we find that the planetary signal is generally too weak to allow for differentiating between the atmospheric models. Only in the case of the H$_2$O versus CO$_2$ models, we find that using MIRI approximately 19 eclipses may do the trick.

\subsubsection{Temperature Inversion}

\begin{figure*}
\begin{center}
\begin{minipage}{0.49\textwidth}
\includegraphics[width=\textwidth]{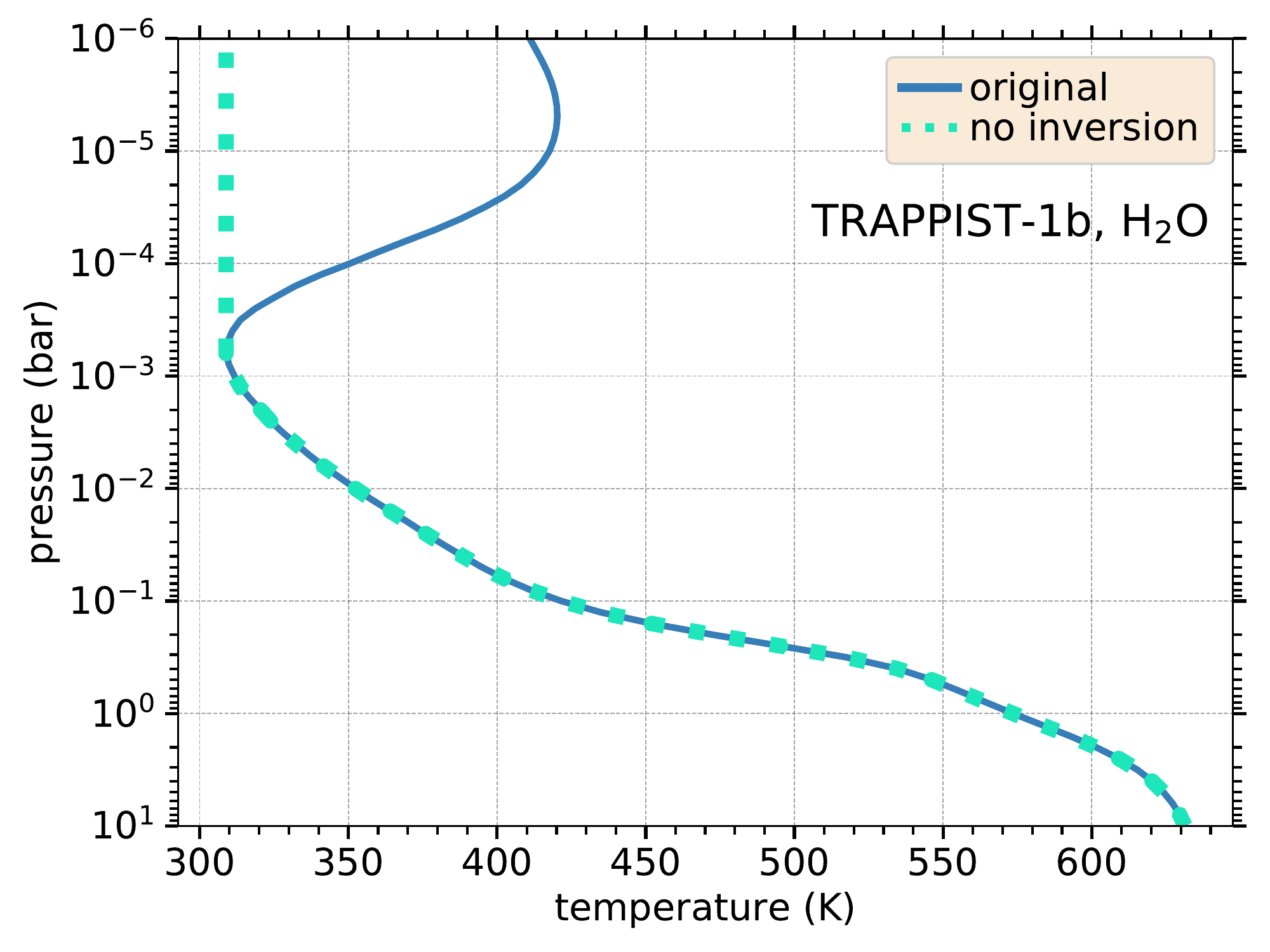}
\end{minipage}
\hfill
\begin{minipage}{0.49\textwidth}
\includegraphics[width=\textwidth]{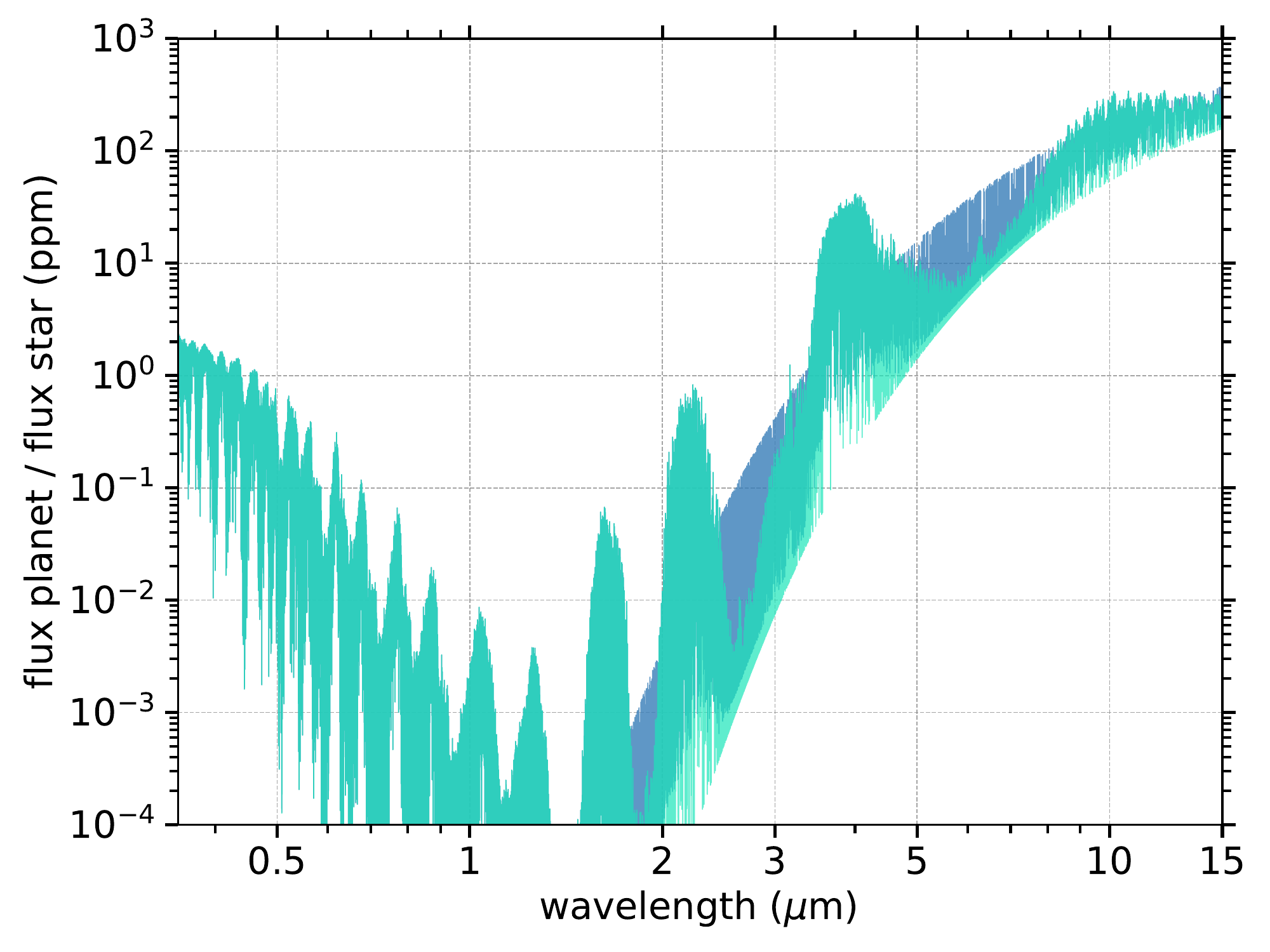}
\end{minipage}
\begin{minipage}{0.49\textwidth}
\includegraphics[width=\textwidth]{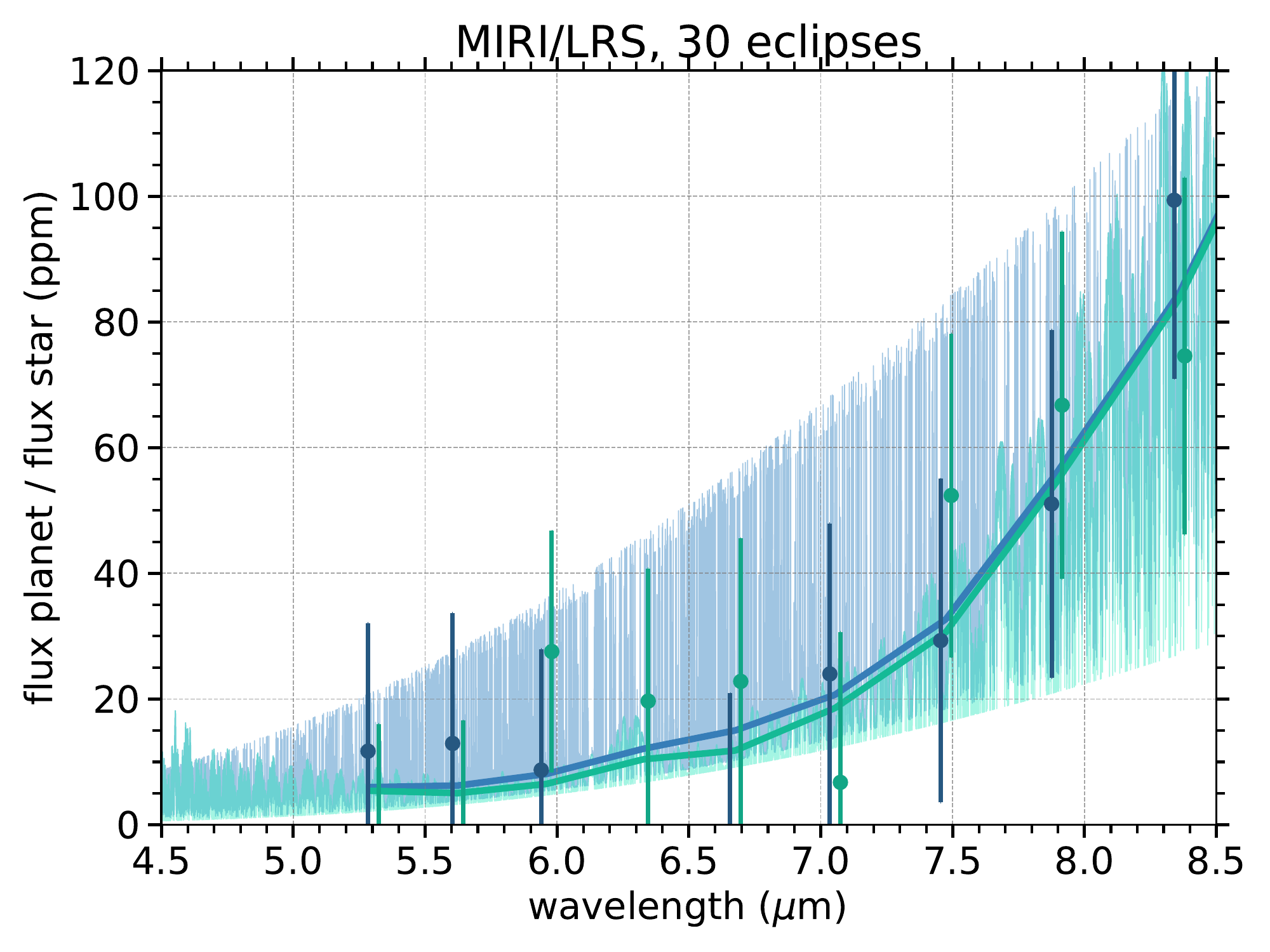}
\end{minipage}
\hfill
\begin{minipage}{0.49\textwidth}
\includegraphics[width=\textwidth]{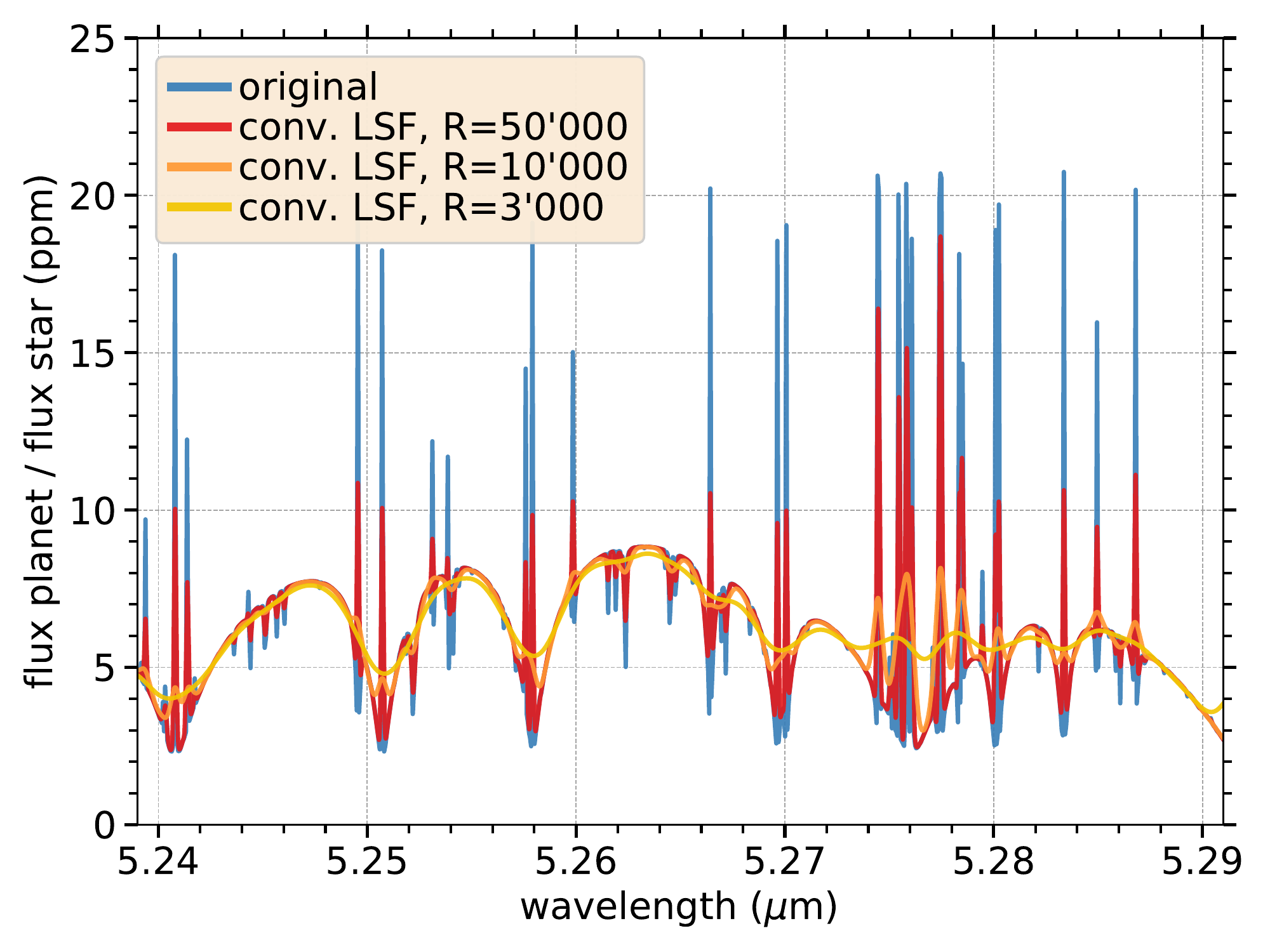}
\end{minipage}
\vspace{-0.3cm}
\caption{{\bf Top left:} TRAPPIST-1b temperature profiles for the pure H$_2$O atmosphere. From the original profile in radiative-convective equilibrium, we artificially construct another test model, which is identical but without the temperature inversion. {\bf Top right:} emission spectra of the models shown on the left. {\bf Bottom left:} simulated JWST secondary eclipse spectra observed with MIRI over 30 eclipses for the two models shown above, rebinned to $R=10$. The theoretical \texttt{HELIOS} spectra are shown rebinned to the mock data (solid lines) and in original resolution (shaded). The data points are slightly offset against each other for clarity. The colors in the top and bottom left panels match. {\bf Bottom right:} zoom of the original \texttt{HELIOS} spectrum calculated here at a resolution of $R=100\,000$ around 5.3 $\mu$m, the longwave limit of CRIRES+. Overlaid are spectra, which are convolved with Gaussians at various resolving powers.}
\label{fig:detect_inv}
\end{center}
\end{figure*}

We focus here on the H$_2$O atmosphere using TRAPPIST-1b parameters, as this scenario provides the most pronounced temperature inversion among the explored models. In the planetary spectrum, the existence of an inversion can be inferred from the presence of emission features in addition to absorption features. The strength of the features hereby correlates with the temperature difference in the inversion layers, similarly to the depth of the absorption features which are indicative of a temperature decline with altitude.

Fig.~\ref{fig:detect_inv}, top left, shows two temperature profiles that we will compare. In addition to our original result, we include a model with an artificial removal of the inversion by extending the temperature from its minimum point upward at a constant value. Thus, the temperature inversion becomes the only difference between these otherwise identical atmospheres. Note that the ``no inversion'' model is not in strict radiative-convective equilibrium and should not be treated as a realistic atmosphere. Specifically, the inversion model adds emission features in wavelengths where the molecular absorption is strongest and the thermal emission has a significant contribution. In this case, those locations are the absorption bands of H$_2$O around 1.9, 2.8 and $5 - 8$ $\mu$m (see Fig.~\ref{fig:detect_inv}, top right panel), with the latter having the larger flux contrast between the inversion and no inversion models. 

A simulated {\it JWST} observation with MIRI LRS at these wavelengths, using the same \texttt{PandExo} parameters as in Sect.~\ref{sec:detect_chem}, suggests that the two models are indistinguishable, independent of the rebinning resolution (see Fig.~\ref{fig:detect_inv}, bottom left panel). The emission features prove to be too narrow to have a significant impact on the overall flux in MIRI's bandpass. In fact, according to our calculation, the TOA flux of the inversion model, integrated over $5 - 8$ $\mu$m, is only $\approx 10\%$ higher than in the other model. Another test using the H$_2$O model of LHS 3844b leads to the same negative result. Although the planetary signal is around three times higher, the inversion is too small to leave a detectable impact on the spectrum (not shown).

While space-borne observations at low resolution do not appear to be the correct tool, high-dispersion spectroscopy may just do the trick. With sufficient resolving power the individual emission lines may be verifiable with the cross-correlation technique applied to ground-based observations \citep[e.g.][]{snellen10, birkby13, dekok13}. For this test we generate the theoretical spectrum at $R=100\,000$. Following the approach described in \citet{pino18a}, we convolve the original spectrum with Gaussians in order to simulate observations at various resolving powers and estimate the required limit to resolve the emission lines. According to our analysis this limit is $R \sim 50\,000$ (see Fig.~\ref{fig:detect_inv}, bottom right panel). This is consistent with the similar previous estimate of \citet{pino18b} (their Fig.~2). Note that the highest resolving power possible with {\it JWST} is $R\sim 3000$, in which case the recorded flux merely assumes the shape of the pressure-broadened line wings.

Unfortunately, in the near future, there will be only one instrument capable of measuring at the required infrared wavelengths, namely CRIRES+ with the upper detector limit at 5.3 $\mu$m \citep{seemann14}. The low signal strength of a terrestrial planet around a faint M dwarf may pose a problem too. In our example the TRAPPIST-1b-like planet exhibits an emission line contrast of 15 ppm (at 5.3 $\mu$m). Together with the star's $K$-band magnitude of 10.3 \citep{gillon17} this goal may be just out of reach of current and near-future instruments. Although there are many high-resolution studies that successfully detected molecular lines in emission spectra, both the planet-to-star contrast and the stellar brightness were always larger than given here (e.g., \citealt{brogi12, birkby13, dekok13}). However, in the mid- to longterm future, a variety of planned high-resolution instruments mounted on 30-meter class telescopes could be ideal for detecting these inversions. For instance, the Extremely Large Telescope's METIS instrument will have the required specifications.

\subsection{Comments on the Modeling Set-up}
\label{sec:param}

Modeling rocky planets with atmospheric temperatures exceeding those of potentially habitable planets comes with two advantages. First, modeling atmospheres above the water condensation regime allows us to generate atmospheres without the necessity to include the intricate gas-liquid phase transition of water, which impacts atmospheric temperatures through latent heat release. Second, condensation results in the formation of clouds, which may mask spectral features, diminishing the information recoverable from these atmospheres (e.g., \citealt{morley13, morley15}). The temperature regime in our case study is too high for thermodynamically formed water clouds or Venus-like sulfuric condensates \citep{buck81, sanchez-lavega04, lincowski18}, but the solar model of LHS 3844b could potentially exhibit KCl condensates. In addition to thermodynamically formed clouds, the comparatively strong UV-flux of M stars may promote the formation of photochemically formed hazes \citep{horst18, fleury19}.

Only extended atmospheres provide the necessary optical thickness to imprint deep spectral features on the planetary emission. Since not all planets are expected to feature substantial envelopes throughout their lifetime, we recommend prior to any in-depth atmospheric characterization attempts to first conduct tests to assess the existence of suitable candidate atmospheres. 

For instance, this can be done via secondary eclipse photometry in the infrared, as explored in our companion study \citet{koll19a}. The gist is that in extended atmospheres zonal winds will transport a fraction of the incoming stellar flux as heat to the night side, thus reducing the observed dayside flux compared to a no-atmosphere case. 

Another strong indication for a thick atmosphere could be the inference of a high planetary Bond albedo, as suggested by our other companion study \citet{mansfield19}. That study argues that there is an upper limit to the surface albedo for any reasonable geochemical composition. Thus, measuring an albedo above this limit would point toward the presence of atmospheric scatterers. 

By assuming an atmospheric pressure of 10 bars, we inherently imply that the planet in question has already passed such a first atmospheric assessment successfully. Our models are thus to be understood under the premise that there is prior knowledge for an extended planetary envelope to be present. 

Our chosen value of 0.1 for the surface albedo is based on solar system findings that rocky bodies without an extended atmosphere and icy surface are dark. For instance, the Bond albedos of Mercury, the Moon and Ceres are all within $0 - 0.2$ \citep{lundock09, madden18}. Also, potential surface materials for exoplanets are thought to possess low albedos \citep{hu12, mansfield19}.

\subsection{Consequences of temperature inversions}
\label{sec:comp}

Since temperature profiles with inversions exhibit cooler lower atmospheres compared to non-inverted profiles, they may activate condensation for species close to their saturation levels, potentially opposing predictions based on non-inverted profiles. Such an enhancement in chemical stratification, via cold trapping condensates at high pressures, could impair characterization efforts with transmission spectroscopy.

In theory, temperatures in the upper atmosphere are the driving factor for Rayleigh-Jeans atmospheric escape so that the prevalent occurrence of temperature inversions in atmospheres of M dwarf exoplanets could lead to higher atmospheric loss rates than previously thought. However, in practice upper limits on the atmospheric mass loss are commonly estimated using energy-limited escape rates, given by the energy of incident XUV photons, so that the Rayleigh-Jeans impact is probably minor \citep{lammer03, selsis07, pierrehumbert10, owen16}.

\subsection{Comparison to other works}
\label{sec:comp}

Recently, \citet{morley17} assessed the feasibility and yield of {\it JWST} observations of the TRAPPIST-1 planets, GJ 1132b, and LHS 1140b. Comparing our CO$_2$ models with their Venus composition models of TRAPPIST-1b and GJ 1132b, we find that our spectra exhibit a significantly lower flux in the spectral windows of CO$_2$, e.g., $3-4$ $\mu$m and $5-7$ $\mu$m. The main cause for this discrepancy is their use of a simplified, analytical temperature profile, which is defined by a constant skin temperature above 0.1 bar and an adiabatic lapse rate below. In contrast, as pointed out in Sect.~\ref{sec:inv}, we find that planets around M stars exhibit smaller temperature gradients in the lower atmosphere, which accordingly mutes the size of the spectral features. This discrepancy applies to both TRAPPIST-1b and GJ 1132b models. We also derive lower atmospheric temperatures because our CO$_2$ atmosphere has a larger Bond albedo. This reduces the size of the predicted thermal signal.

As part of their habitability study of the TRAPPIST-1 system, \citet{lincowski18} generated a cloud-free TRAPPIST-1b atmospheric model with a Venus-like composition and 10 bar surface pressure, similar to our CO$_2$ case. They too modeled the atmosphere in radiative-convective equilibrium and found a temperature profile qualitatively similar to ours. Theirs and our temperature profiles fully coincide in the intermediate atmospheric layers, 0.1 bar $< P < 1$ mbar, and exhibit the same inversion trend in the upper layers, $P < 1$ mbar, with our model being only slightly cooler there. The largest difference is found in the lower atmosphere, $P > 0.1$ bar, where they predict a steeper lapse rate and higher temperatures close to the surface. Consequently, they predict a somewhat larger thermal emission in the CO$_2$ opacity windows. This temperature discrepancy is caused by the addition of H$_2$O in their Venus-like atmosphere. Since H$_2$O features less pronounced opacity windows in the mid-infrared than CO$_2$, the atmosphere can cool less efficiently, which drives the temperatures up. By adding H$_2$O to our CO$_2$ model we can qualitatively reproduce their results (not shown).

Exploring the climate of Proxima Centauri b, \citet{meadows18} also obtained temperature inversions in the upper atmosphere, $P < 1$ mbar, in their CO$_2$-rich model. The parameters of this system, ($R_{\rm star} / a = 0.014$) and $T_{\rm star} = 3050$ K \citep{anglada16}, place this planet deep into the inversion regime in Fig. \ref{fig:inv_contour}, bottom left panel, consistent with our prediction on the emergence of temperature inversions.

Lastly, we compare with \citet{lustig-yaeger19}, who conducted a thorough observability study of the TRAPPIST-1 system for JWST. They found that for TRAPPIST-1b a 10 bar clear-sky CO$_2$ atmosphere could be detected with MIRI LRS if observed for 20-30 eclipses. This is broadly consistent with our estimates, which range between 19 to more than 30 eclipses necessary to distinguish the CO$_2$ atmosphere from the H$_2$O or solar cases with this instrument.

\section{Conclusions}
\label{sec:concl}

The principal goal of the current study is to better understand the radiative implications of M stars on atmospheric temperatures of rocky exoplanets with a surface. To this end, we have presented self-consistently calculated models of atmospheres in radiative-convective equilibrium, taking TRAPPIST-1b, GJ 1132b and LHS 3844b as case studies. Using  three  contrasting  atmospheric  compositions $-$ pure H$_2$O, pure CO$_2$, and hydrogen-dominated with solar abundances $-$ our results indicate the following salient characteristics.

\begin{itemize}\itemsep1.5pt
\item{The peak emission of M stars in the near-infrared enables the greenhouse gases H$_2$O and CO$_2$ to act as shortwave absorbers, causing a gradual attenuation of the incoming stellar flux throughout the atmosphere. This leads to a reduction in the amount of energy deposited in the lower atmospheric layers, effectively suppressing convection if the atmosphere is thick enough. Furthermore, a temperature inversion appears if the gas in the upper atmosphere cannot efficiently cool.}
\item{The emergence of temperature inversions depends mainly on the planet's equilibrium temperature and only secondarily on the stellar temperature and spectral energy distribution.  With increasing temperature, the opacity windows of the absorbing gases are filled up, which enhances the upper atmosphere's ability to cool, acting against the creation of an inversion. Therefore, we predict temperature inversions to be common in the atmospheres of planets orbiting M stars, if the atmospheric temperatures are sufficiently low. Conversely, the occurrence of inversions is diminished or even fully suppressed as temperatures rise. Interestingly, this behavior contrasts the emergence of temperature inversions in hot Jupiters, where the only the very hot atmospheres are expected to feature inversions.}
\item{The CO$_2$ atmospheres feature overall lower temperatures and stronger reflectance spectra than the other two composition cases. Since the CO$_2$ opacity in the near-infrared is comparatively weak, a significant fraction of radiation is reflected back by Rayleigh scattering and the surface instead of being absorbed by the atmosphere. This is reflected by the much higher Bond albedo $A_{\rm B} \approx 0.3$ of the CO$_2$ model, compared to the other cases, where $A_{\rm B} \lesssim 0.01$. This effect is expected to diminish if additional greenhouse gases such as water vapor are added in modest mixing ratios, which will act to fill in the substantial opacity windows.}
\end{itemize}

Regarding observability, we estimate that for LHS 3844b two eclipse runs with {\it JWST} should be sufficient to distinguish between the different atmospheric models as defined in this work. The fainter GJ 1132b requires for the same task $\sim 5 - 15$ eclipse observations and for TRAPPIST-1b it is only possible to distinguish between the H$_2$O and CO$_2$ models with approximately 19 eclipses. The emission features indicative of the temperature inversions are not resolvable by {\it JWST}. Ground-based high-dispersion instruments have the necessary resolution, but the weak signal may prove problematic.

We expect the herein explored models to form a representative basis of cornerstone atmospheres sampling the possible parameter space of not yet well-understood atmospheric properties of rocky exoplanets. In future works, we plan to build on this foundation and include additions increasing the sophistication and physical realism of the atmospheric modeling.

\section*{Acknowledgements}

M.Malik acknowledges support from the Swiss National Science Foundation under the Early Postdoc Mobility grant P2BEP2\_181705. E.M.-R.Kempton acknowledges support from the National Science Foundation under Grant No.1654295 and from the Research Corporation for Science Advancement through their Cottrell Scholar program. D.D.B.Koll was supported by a James McDonnell Foundation postdoctoral fellowship. J.L.Bean acknowledges support from the David and Lucile Packard Foundation. M.Mansfield and E.Kite were supported by NASA grant NNX16AB44G.

\software{\texttt{HELIOS} \citep{malik17, malik19}, \texttt{HELIOS-K} \citep{grimm15}, \texttt{FastChem} \citep{stock18}, \texttt{CUDA} \citep{cuda}, \texttt{PyCUDA} \citep{pycuda}, \texttt{python} \citep{python}, \texttt{scipy} \citep{scipy}, \texttt{numpy} \citep{numpy}, \texttt{matplotlib} \citep{matplotlib}}

\appendix
\vspace{-0.1cm}
\section{Implementation of a surface layer}
\label{app:surface}

For this work, we have extended HELIOS to include the a solid surface layer as a lower boundary to the atmosphere. The surface participates in the radiative convergence iteration via its temperature until it is in radiative equilibrium with the overlying atmosphere. The surface temperature is calculated as
\begin{equation}
T_{\rm surf} = \left[\frac{(1-A)\mathcal{F}_{\downarrow, \rm BOA}}{\epsilon\sigma_{\rm SB}} + \frac{T_{\rm interior}^4}{\epsilon}\right]^{1/4} ,   
\end{equation}
where $A$ is the surface albedo, $\mathcal{F}_{\downarrow, \rm BOA}$ is the downward flux at the bottom-of-atmosphere (BOA), $\epsilon \equiv (1-A)$ is the surface emissivity, $\sigma_{\rm SB}$ is the Stefan-Boltzmann constant and $T_{\rm interior}$ is the interior temperature of the planet, a surrogate for the interior heat flux $\mathcal{F}_{\rm interior} = \sigma_{\rm SB} T_{\rm interior}^4$. Note, that $\mathcal{F}$ represents a flux integrated over all wavelengths. The upward flux at BOA reads
\begin{equation}
\mathcal{F}_{\uparrow, {\rm BOA}} = A \mathcal{F}_{\downarrow, {\rm BOA}} + \epsilon\sigma_{\rm SB} T_{\rm surf}^4 ,
\end{equation}
conserving the boundary condition $\mathcal{F}_{\uparrow, {\rm BOA}}  = \mathcal{F}_{\downarrow, {\rm BOA}} + \mathcal{F}_{\rm interior}$, independent of the surface albedo. Numerically, the surface acts as a ``ghost layer'' at BOA, by taking both the downward atmospheric flux and the interior heat flux into account.

One complication is that due to the limited wavelength coverage of the numerical model, integrating the thermal flux over all wavelengths will always return less than the correct value from the Stefan-Boltzmann law. Hence, the thermal emission must be adjusted as
\begin{equation}
\label{eq:Bcorr}
B_{\lambda, \rm corr}(T_{\rm surf}) = B_{\lambda, \rm uncorr}(T_{\rm surf})\frac{\sigma_{\rm SB}T_{\rm surf}^4}{\pi \int_{\lambda_{\rm min}}^{\lambda_{\rm max}} B_{\lambda, \rm uncorr}(T_{\rm surf})~d\lambda} ,
\end{equation}
where $B_{\lambda, \rm corr}$ is the corrected blackbody function for $T_{\rm surf}$ at wavelength $\lambda$, $B_{\lambda, \rm uncorr}$ is the analogous uncorrected blackbody function, and $\lambda_{\rm min}$ and $\lambda_{\rm max}$ are the lower and upper wavelength boundaries of the model. As the fraction term in eq.~(\ref{eq:Bcorr}) is always larger than unity, not correcting for this discrepancy results in an artificial energy sink at the surface. Note that the emissivity should be present in both the numerator and denominator terms, but for a wavelength-independent emissivity they cancel each other out. The correct upward spectral flux at BOA is given by
\begin{equation}
F_{\uparrow, {\rm BOA}} = A F_{\downarrow, {\rm BOA}} + \epsilon\pi B_{\lambda, \rm corr}(T_{\rm surf}) ,
\end{equation}
where $F$ represents a wavelength-dependent (per unit wavelength) flux. 

Note, that this formulation holds only for hot and dry atmospheres, beyond temperatures where condensation is expected. For cooler planets, the latent and sensible heat need to be added to the near-surface fluxes.

\bibliographystyle{aasjournal}

\bibliography{mybib}

\begin{thebibliography}{}
\expandafter\ifx\csname natexlab\endcsname\relax\def\natexlab#1{#1}\fi
\providecommand{\url}[1]{\href{#1}{#1}}
\providecommand{\dodoi}[1]{doi:~\href{http://doi.org/#1}{\nolinkurl{#1}}}
\providecommand{\doeprint}[1]{\href{http://ascl.net/#1}{\nolinkurl{http://ascl.net/#1}}}
\providecommand{\doarXiv}[1]{\href{https://arxiv.org/abs/#1}{\nolinkurl{https://arxiv.org/abs/#1}}}

\bibitem[{{Anglada-Escud{\'e}} {et~al.}(2016){Anglada-Escud{\'e}}, {Amado},
  {Barnes}, {Berdi{\~n}as}, {Butler}, {Coleman}, {de La Cueva}, {Dreizler},
  {Endl}, {Giesers}, {Jeffers}, {Jenkins}, {Jones}, {Kiraga}, {K{\"u}rster},
  {L{\'o}pez-Gonz{\'a}lez}, {Marvin}, {Morales}, {Morin}, {Nelson}, {Ortiz},
  {Ofir}, {Paardekooper}, {Reiners}, {Rodr{\'{\i}}guez},
  {Rodr{\'{\i}}guez-L{\'o}pez}, {Sarmiento}, {Strachan}, {Tsapras}, {Tuomi}, \&
  {Zechmeister}}]{anglada16}
{Anglada-Escud{\'e}}, G., {Amado}, P.~J., {Barnes}, J., {et~al.} 2016, \nat,
  536, 437, \dodoi{10.1038/nature19106}

\bibitem[{{Arcangeli} {et~al.}(2018){Arcangeli}, {D{\'e}sert}, {Line}, {Bean},
  {Parmentier}, {Stevenson}, {Kreidberg}, {Fortney}, {Mansfield}, \&
  {Showman}}]{arcangeli18}
{Arcangeli}, J., {D{\'e}sert}, J.-M., {Line}, M.~R., {et~al.} 2018, \apj, 855,
  L30, \dodoi{10.3847/2041-8213/aab272}

\bibitem[{{Asplund} {et~al.}(2009){Asplund}, {Grevesse}, {Sauval}, \&
  {Scott}}]{asplund09}
{Asplund}, M., {Grevesse}, N., {Sauval}, A.~J., \& {Scott}, P. 2009, \araa, 47,
  481, \dodoi{10.1146/annurev.astro.46.060407.145222}

\bibitem[{{Azzam} {et~al.}(2016){Azzam}, {Tennyson}, {Yurchenko}, \&
  {Naumenko}}]{azzam16}
{Azzam}, A.~A.~A., {Tennyson}, J., {Yurchenko}, S.~N., \& {Naumenko}, O.~V.
  2016, \mnras, 460, 4063, \dodoi{10.1093/mnras/stw1133}

\bibitem[{{Barber} {et~al.}(2006){Barber}, {Tennyson}, {Harris}, \&
  {Tolchenov}}]{barber06}
{Barber}, R.~J., {Tennyson}, J., {Harris}, G.~J., \& {Tolchenov}, R.~N. 2006,
  \mnras, 368, 1087, \dodoi{10.1111/j.1365-2966.2006.10184.x}

\bibitem[{{Barstow} \& {Irwin}(2016)}]{barstow16}
{Barstow}, J.~K., \& {Irwin}, P.~G.~J. 2016, \mnras, 461, L92,
  \dodoi{10.1093/mnrasl/slw109}

\bibitem[{Batalha {et~al.}(2018)Batalha, Lewis, Line, Valenti, \&
  Stevenson}]{batalha18}
Batalha, N.~E., Lewis, N.~K., Line, M.~R., Valenti, J., \& Stevenson, K. 2018,
  The Astrophysical Journal, 856, L34, \dodoi{10.3847/2041-8213/aab896}

\bibitem[{{Batalha} {et~al.}(2017){Batalha}, {Mandell}, {Pontoppidan},
  {Stevenson}, {Lewis}, {Kalirai}, {Earl}, {Greene}, {Albert}, \&
  {Nielsen}}]{batalha17a}
{Batalha}, N.~E., {Mandell}, A., {Pontoppidan}, K., {et~al.} 2017, \pasp, 129,
  064501, \dodoi{10.1088/1538-3873/aa65b0}

\bibitem[{{Berta-Thompson} {et~al.}(2015){Berta-Thompson}, {Irwin},
  {Charbonneau}, {Newton}, {Dittmann}, {Astudillo-Defru}, {Bonfils}, {Gillon},
  {Jehin}, {Stark}, {Stalder}, {Bouchy}, {Delfosse}, {Forveille}, {Lovis},
  {Mayor}, {Neves}, {Pepe}, {Santos}, {Udry}, \& {W{\"u}nsche}}]{berta15}
{Berta-Thompson}, Z.~K., {Irwin}, J., {Charbonneau}, D., {et~al.} 2015, \nat,
  527, 204, \dodoi{10.1038/nature15762}

\bibitem[{{Birkby} {et~al.}(2013){Birkby}, {de Kok}, {Brogi}, {de Mooij},
  {Schwarz}, {Albrecht}, \& {Snellen}}]{birkby13}
{Birkby}, J.~L., {de Kok}, R.~J., {Brogi}, M., {et~al.} 2013, \mnras, 436, L35,
  \dodoi{10.1093/mnrasl/slt107}

\bibitem[{{Bolmont} {et~al.}(2017){Bolmont}, {Selsis}, {Owen}, {Ribas},
  {Raymond}, {Leconte}, \& {Gillon}}]{bolmont17}
{Bolmont}, E., {Selsis}, F., {Owen}, J.~E., {et~al.} 2017, \mnras, 464, 3728,
  \dodoi{10.1093/mnras/stw2578}

\bibitem[{{Bonfils} {et~al.}(2018){Bonfils}, {Almenara}, {Cloutier},
  {W{\"u}nsche}, {Astudillo-Defru}, {Berta-Thompson}, {Bouchy}, {Charbonneau},
  {Delfosse}, {D{\'\i}az}, {Dittmann}, {Doyon}, {Forveille}, {Irwin}, {Lovis},
  {Mayor}, {Menou}, {Murgas}, {Newton}, {Pepe}, {Santos}, \&
  {Udry}}]{bonfils18}
{Bonfils}, X., {Almenara}, J.~M., {Cloutier}, R., {et~al.} 2018, \aap, 618,
  A142, \dodoi{10.1051/0004-6361/201731884}

\bibitem[{{Brogi} {et~al.}(2012){Brogi}, {Snellen}, {de Kok}, {Albrecht},
  {Birkby}, \& {de Mooij}}]{brogi12}
{Brogi}, M., {Snellen}, I.~A.~G., {de Kok}, R.~J., {et~al.} 2012, \nat, 486,
  502, \dodoi{10.1038/nature11161}

\bibitem[{Buck(1981)}]{buck81}
Buck, A.~L. 1981, Journal of Applied Meteorology, 20, 1527,
  \dodoi{10.1175/1520-0450(1981)020<1527:NEFCVP>2.0.CO;2}

\bibitem[{{Burrows} {et~al.}(2008){Burrows}, {Budaj}, \& {Hubeny}}]{burrows08}
{Burrows}, A., {Budaj}, J., \& {Hubeny}, I. 2008, \apj, 678, 1436,
  \dodoi{10.1086/533518}

\bibitem[{{Burrows} {et~al.}(2000){Burrows}, {Marley}, \& {Sharp}}]{burrows00}
{Burrows}, A., {Marley}, M.~S., \& {Sharp}, C.~M. 2000, \apj, 531, 438,
  \dodoi{10.1086/308462}

\bibitem[{{Burrows} \& {Volobuyev}(2003)}]{burrows03}
{Burrows}, A., \& {Volobuyev}, M. 2003, \apj, 583, 985, \dodoi{10.1086/345412}

\bibitem[{{Chen} \& {Rogers}(2016)}]{chen16}
{Chen}, H., \& {Rogers}, L.~A. 2016, \apj, 831, 180,
  \dodoi{10.3847/0004-637X/831/2/180}

\bibitem[{{Chen} \& {Kipping}(2017)}]{chen17}
{Chen}, J., \& {Kipping}, D. 2017, \apj, 834, 17,
  \dodoi{10.3847/1538-4357/834/1/17}

\bibitem[{{Cowan} {et~al.}(2015){Cowan}, {Greene}, {Angerhausen}, {Batalha},
  {Clampin}, {Col{\'o}n}, {Crossfield}, {Fortney}, {Gaudi}, {Harrington},
  {Iro}, {Lillie}, {Linsky}, {Lopez-Morales}, {Mandell}, \&
  {Stevenson}}]{cowan15}
{Cowan}, N.~B., {Greene}, T., {Angerhausen}, D., {et~al.} 2015, \pasp, 127,
  311, \dodoi{10.1086/680855}

\bibitem[{{Cox}(2000)}]{cox00}
{Cox}, A.~N. 2000, {Allen's astrophysical quantities}

\bibitem[{{de Kok} {et~al.}(2013){de Kok}, {Brogi}, {Snellen}, {Birkby},
  {Albrecht}, \& {de Mooij}}]{dekok13}
{de Kok}, R.~J., {Brogi}, M., {Snellen}, I.~A.~G., {et~al.} 2013, \aap, 554,
  A82, \dodoi{10.1051/0004-6361/201321381}

\bibitem[{{de Wit} {et~al.}(2016){de Wit}, {Wakeford}, {Gillon}, {Lewis},
  {Valenti}, {Demory}, {Burgasser}, {Burdanov}, {Delrez}, {Jehin}, {Lederer},
  {Queloz}, {Triaud}, \& {Van Grootel}}]{dewit16}
{de Wit}, J., {Wakeford}, H.~R., {Gillon}, M., {et~al.} 2016, \nat, 537, 69,
  \dodoi{10.1038/nature18641}

\bibitem[{{Delrez} {et~al.}(2018){Delrez}, {Gillon}, {Triaud}, {Demory}, {de
  Wit}, {Ingalls}, {Agol}, {Bolmont}, {Burdanov}, {Burgasser}, {Carey},
  {Jehin}, {Leconte}, {Lederer}, {Queloz}, {Selsis}, \& {Van
  Grootel}}]{delrez18}
{Delrez}, L., {Gillon}, M., {Triaud}, A.~H.~M.~J., {et~al.} 2018, \mnras, 475,
  3577, \dodoi{10.1093/mnras/sty051}

\bibitem[{{Diamond-Lowe} {et~al.}(2018){Diamond-Lowe}, {Berta-Thompson},
  {Charbonneau}, \& {Kempton}}]{diamond18}
{Diamond-Lowe}, H., {Berta-Thompson}, Z., {Charbonneau}, D., \& {Kempton},
  E.~M.-R. 2018, \aj, 156, 42, \dodoi{10.3847/1538-3881/aac6dd}

\bibitem[{Dong {et~al.}(2018)Dong, Jin, Lingam, Airapetian, Ma, \& van~der
  Holst}]{dong18}
Dong, C., Jin, M., Lingam, M., {et~al.} 2018, Proceedings of the National
  Academy of Sciences, 115, 260, \dodoi{10.1073/pnas.1708010115}

\bibitem[{{Dressing} \& {Charbonneau}(2015)}]{dressing15}
{Dressing}, C.~D., \& {Charbonneau}, D. 2015, \apj, 807, 45,
  \dodoi{10.1088/0004-637X/807/1/45}

\bibitem[{{Elkins-Tanton} \& {Seager}(2008)}]{elkins08}
{Elkins-Tanton}, L.~T., \& {Seager}, S. 2008, \apj, 685, 1237,
  \dodoi{10.1086/591433}

\bibitem[{{Fleury} {et~al.}(2019){Fleury}, {Gudipati}, {Henderson}, \&
  {Swain}}]{fleury19}
{Fleury}, B., {Gudipati}, M.~S., {Henderson}, B.~L., \& {Swain}, M. 2019, \apj,
  871, 158, \dodoi{10.3847/1538-4357/aaf79f}

\bibitem[{{Fortney} {et~al.}(2008){Fortney}, {Lodders}, {Marley}, \&
  {Freedman}}]{fortney08}
{Fortney}, J.~J., {Lodders}, K., {Marley}, M.~S., \& {Freedman}, R.~S. 2008,
  \apj, 678, 1419, \dodoi{10.1086/528370}

\bibitem[{{Fulton} {et~al.}(2017){Fulton}, {Petigura}, {Howard}, {Isaacson},
  {Marcy}, {Cargile}, {Hebb}, {Weiss}, {Johnson}, {Morton}, {Sinukoff},
  {Crossfield}, \& {Hirsch}}]{fulton17}
{Fulton}, B.~J., {Petigura}, E.~A., {Howard}, A.~W., {et~al.} 2017, \aj, 154,
  109, \dodoi{10.3847/1538-3881/aa80eb}

\bibitem[{{Gaidos} {et~al.}(2016){Gaidos}, {Mann}, {Kraus}, \&
  {Ireland}}]{gaidos16}
{Gaidos}, E., {Mann}, A.~W., {Kraus}, A.~L., \& {Ireland}, M. 2016, \mnras,
  457, 2877, \dodoi{10.1093/mnras/stw097}

\bibitem[{{Gillon} {et~al.}(2017){Gillon}, {Triaud}, {Demory}, {Jehin}, {Agol},
  {Deck}, {Lederer}, {de Wit}, {Burdanov}, {Ingalls}, {Bolmont}, {Leconte},
  {Raymond}, {Selsis}, {Turbet}, {Barkaoui}, {Burgasser}, {Burleigh}, {Carey},
  {Chaushev}, {Copperwheat}, {Delrez}, {Fernandes}, {Holdsworth}, {Kotze}, {Van
  Grootel}, {Almleaky}, {Benkhaldoun}, {Magain}, \& {Queloz}}]{gillon17}
{Gillon}, M., {Triaud}, A.~H.~M.~J., {Demory}, B.-O., {et~al.} 2017, \nat, 542,
  456, \dodoi{10.1038/nature21360}

\bibitem[{Gordon {et~al.}(2017)Gordon, Rothman, Hill, Kochanov, Tan, Bernath,
  Birk, Boudon, Campargue, Chance, Drouin, Flaud, Gamache, Hodges, Jacquemart,
  Perevalov, Perrin, Shine, Smith, Tennyson, Toon, Tran, Tyuterev, Barbe,
  Császár, Devi, Furtenbacher, Harrison, Hartmann, Jolly, Johnson, Karman,
  Kleiner, Kyuberis, Loos, Lyulin, Massie, Mikhailenko, Moazzen-Ahmadi,
  Müller, Naumenko, Nikitin, Polyansky, Rey, Rotger, Sharpe, Sung, Starikova,
  Tashkun, Auwera, Wagner, Wilzewski, Wcisło, Yu, \& Zak}]{gordon17}
Gordon, I., Rothman, L., Hill, C., {et~al.} 2017, Journal of Quantitative
  Spectroscopy and Radiative Transfer, 203, 3 ,
  \dodoi{https://doi.org/10.1016/j.jqsrt.2017.06.038}

\bibitem[{{Grimm} \& {Heng}(2015)}]{grimm15}
{Grimm}, S.~L., \& {Heng}, K. 2015, \apj, 808, 182,
  \dodoi{10.1088/0004-637X/808/2/182}

\bibitem[{{Grimm} {et~al.}(2018){Grimm}, {Demory}, {Gillon}, {Dorn}, {Agol},
  {Burdanov}, {Delrez}, {Sestovic}, {Triaud}, {Turbet}, {Bolmont}, {Caldas},
  {Wit}, {Jehin}, {Leconte}, {Raymond}, {Grootel}, {Burgasser}, {Carey},
  {Fabrycky}, {Heng}, {Hernandez}, {Ingalls}, {Lederer}, {Selsis}, \&
  {Queloz}}]{grimm18}
{Grimm}, S.~L., {Demory}, B.-O., {Gillon}, M., {et~al.} 2018, \aap, 613, A68,
  \dodoi{10.1051/0004-6361/201732233}

\bibitem[{{Gully-Santiago} {et~al.}(2017){Gully-Santiago}, {Herczeg},
  {Czekala}, {Somers}, {Grankin}, {Covey}, {Donati}, {Alencar}, {Hussain},
  {Shappee}, {Mace}, {Lee}, {Holoien}, {Jose}, \& {Liu}}]{gully-santiago17}
{Gully-Santiago}, M.~A., {Herczeg}, G.~J., {Czekala}, I., {et~al.} 2017, \apj,
  836, 200, \dodoi{10.3847/1538-4357/836/2/200}

\bibitem[{{Harris} {et~al.}(2006){Harris}, {Tennyson}, {Kaminsky}, {Pavlenko},
  \& {Jones}}]{harris06}
{Harris}, G.~J., {Tennyson}, J., {Kaminsky}, B.~M., {Pavlenko}, Y.~V., \&
  {Jones}, H.~R.~A. 2006, \mnras, 367, 400,
  \dodoi{10.1111/j.1365-2966.2005.09960.x}

\bibitem[{{Heng} {et~al.}(2018){Heng}, {Malik}, \& {Kitzmann}}]{heng18}
{Heng}, K., {Malik}, M., \& {Kitzmann}, D. 2018, \apjs, 237, 29,
  \dodoi{10.3847/1538-4365/aad199}

\bibitem[{{H{\"o}rst} {et~al.}(2018){H{\"o}rst}, {He}, {Lewis}, {Kempton},
  {Marley}, {Morley}, {Moses}, {Valenti}, \& {Vuitton}}]{horst18}
{H{\"o}rst}, S.~M., {He}, C., {Lewis}, N.~K., {et~al.} 2018, Nature Astronomy,
  2, 303, \dodoi{10.1038/s41550-018-0397-0}

\bibitem[{Hu {et~al.}(2012)Hu, Ehlmann, \& Seager}]{hu12}
Hu, R., Ehlmann, B.~L., \& Seager, S. 2012, The Astrophysical Journal, 752, 7,
  \dodoi{10.1088/0004-637x/752/1/7}

\bibitem[{Hunter(2007)}]{matplotlib}
Hunter, J.~D. 2007, Computing In Science \& Engineering, 9, 90,
  \dodoi{10.1109/MCSE.2007.55}

\bibitem[{{Husser} {et~al.}(2013){Husser}, {Wende-von Berg}, {Dreizler},
  {Homeier}, {Reiners}, {Barman}, \& {Hauschildt}}]{husser13}
{Husser}, T.-O., {Wende-von Berg}, S., {Dreizler}, S., {et~al.} 2013, \aap,
  553, A6, \dodoi{10.1051/0004-6361/201219058}

\bibitem[{{Irwin}(2009)}]{irwin09}
{Irwin}, P.~G.~J. 2009, {Giant Planets of Our Solar System} (Berlin: Springer),
  \dodoi{10.1007/978-3-540-85158-5}

\bibitem[{{Jin} {et~al.}(2014){Jin}, {Mordasini}, {Parmentier}, {van Boekel},
  {Henning}, \& {Ji}}]{jin14}
{Jin}, S., {Mordasini}, C., {Parmentier}, V., {et~al.} 2014, \apj, 795, 65,
  \dodoi{10.1088/0004-637X/795/1/65}

\bibitem[{{Kempton} {et~al.}(2018){Kempton}, {Bean}, {Louie}, {Deming}, {Koll},
  {Mansfield}, {Christiansen}, {L{\'o}pez-Morales}, {Swain}, {Zellem},
  {Ballard}, {Barclay}, {Barstow}, {Batalha}, {Beatty}, {Berta-Thompson},
  {Birkby}, {Buchhave}, {Charbonneau}, {Cowan}, {Crossfield}, {de Val-Borro},
  {Doyon}, {Dragomir}, {Gaidos}, {Heng}, {Hu}, {Kane}, {Kreidberg}, {Mallonn},
  {Morley}, {Narita}, {Nascimbeni}, {Pall{\'e}}, {Quintana}, {Rauscher},
  {Seager}, {Shkolnik}, {Sing}, {Sozzetti}, {Stassun}, {Valenti}, \& {von
  Essen}}]{kempton18}
{Kempton}, E.~M.-R., {Bean}, J.~L., {Louie}, D.~R., {et~al.} 2018, \pasp, 130,
  114401, \dodoi{10.1088/1538-3873/aadf6f}

\bibitem[{{Kite} {et~al.}(2009){Kite}, {Manga}, \& {Gaidos}}]{kite09}
{Kite}, E.~S., {Manga}, M., \& {Gaidos}, E. 2009, \apj, 700, 1732,
  \dodoi{10.1088/0004-637X/700/2/1732}

\bibitem[{{Kl{\"o}ckner} {et~al.}(2012){Kl{\"o}ckner}, {Pinto}, {Lee},
  {Catanzaro}, {Ivanov}, \& {Fasih}}]{pycuda}
{Kl{\"o}ckner}, A., {Pinto}, N., {Lee}, Y., {et~al.} 2012, Parallel Computing,
  38, 157, \dodoi{10.1016/j.parco.2011.09.001}

\bibitem[{{Koll}(2019)}]{koll19b}
{Koll}, D. D.~B. 2019, arXiv e-prints, arXiv:1907.13145.
\newblock \doarXiv{1907.13145}

\bibitem[{{Koll} {et~al.}(2019){Koll}, {Malik}, {Mansfield}, {Kempton}, {Kite},
  {Abbot}, \& {Bean}}]{koll19a}
{Koll}, D. D.~B., {Malik}, M., {Mansfield}, M., {et~al.} 2019, arXiv e-prints,
  arXiv:1907.13138.
\newblock \doarXiv{1907.13138}

\bibitem[{{Kopparapu} {et~al.}(2016){Kopparapu}, {Wolf}, {Haqq-Misra}, {Yang},
  {Kasting}, {Meadows}, {Terrien}, \& {Mahadevan}}]{kopparapu16}
{Kopparapu}, R.~k., {Wolf}, E.~T., {Haqq-Misra}, J., {et~al.} 2016, \apj, 819,
  84, \dodoi{10.3847/0004-637X/819/1/84}

\bibitem[{{Kopparapu} {et~al.}(2013){Kopparapu}, {Ramirez}, {Kasting}, {Eymet},
  {Robinson}, {Mahadevan}, {Terrien}, {Domagal-Goldman}, {Meadows}, \&
  {Deshpande}}]{kopparapu13}
{Kopparapu}, R.~K., {Ramirez}, R., {Kasting}, J.~F., {et~al.} 2013, \apj, 765,
  131, \dodoi{10.1088/0004-637X/765/2/131}

\bibitem[{{Kreidberg} {et~al.}(2019){Kreidberg}, {Koll}, {Morley}, {Hu},
  {Schaefer}, {Deming}, {Stevenson}, {Dittmann}, {Vanderburg}, {Berardo},
  {Guo}, {Stassun}, {Crossfield}, {Charbonneau}, {Latham}, {Loeb}, {Ricker},
  {Seager}, \& {Vand erspek}}]{kreidberg19}
{Kreidberg}, L., {Koll}, D. D.~B., {Morley}, C., {et~al.} 2019, arXiv e-prints,
  arXiv:1908.06834.
\newblock \doarXiv{1908.06834}

\bibitem[{{Kurucz}(2011)}]{kurucz11}
{Kurucz}, R.~L. 2011, Canadian Journal of Physics, 89, 417,
  \dodoi{10.1139/p10-104}

\bibitem[{{Lammer} {et~al.}(2003){Lammer}, {Selsis}, {Ribas}, {Guinan},
  {Bauer}, \& {Weiss}}]{lammer03}
{Lammer}, H., {Selsis}, F., {Ribas}, I., {et~al.} 2003, \apjl, 598, L121,
  \dodoi{10.1086/380815}

\bibitem[{{Lee} \& {Kim}(2004)}]{lee04}
{Lee}, H.-W., \& {Kim}, H.~I. 2004, \mnras, 347, 802,
  \dodoi{10.1111/j.1365-2966.2004.07255.x}

\bibitem[{Li {et~al.}(2015)Li, Gordon, Rothman, Tan, Hu, Kassi, Campargue, \&
  Medvedev}]{li15}
Li, G., Gordon, I.~E., Rothman, L.~S., {et~al.} 2015, The Astrophysical Journal
  Supplement Series, 216, 15

\bibitem[{Lincowski {et~al.}(2018)Lincowski, Meadows, Crisp, Robinson, Luger,
  Lustig-Yaeger, \& Arney}]{lincowski18}
Lincowski, A.~P., Meadows, V.~S., Crisp, D., {et~al.} 2018, The Astrophysical
  Journal, 867, 76, \dodoi{10.3847/1538-4357/aae36a}

\bibitem[{{Lopez} {et~al.}(2012){Lopez}, {Fortney}, \& {Miller}}]{lopez12}
{Lopez}, E.~D., {Fortney}, J.~J., \& {Miller}, N. 2012, \apj, 761, 59,
  \dodoi{10.1088/0004-637X/761/1/59}

\bibitem[{{Lopez} \& {Rice}(2018)}]{lopez18}
{Lopez}, E.~D., \& {Rice}, K. 2018, \mnras, 479, 5303,
  \dodoi{10.1093/mnras/sty1707}

\bibitem[{{Lundock} {et~al.}(2009){Lundock}, {Ichikawa}, {Okita}, {Kurita},
  {Kawabata}, {Uemura}, {Yamashita}, {Ohsugi}, {Sato}, \& {Kino}}]{lundock09}
{Lundock}, R., {Ichikawa}, T., {Okita}, H., {et~al.} 2009, \aap, 507, 1649,
  \dodoi{10.1051/0004-6361/200912581}

\bibitem[{{Lustig-Yaeger} {et~al.}(2019){Lustig-Yaeger}, {Meadows}, \&
  {Lincowski}}]{lustig-yaeger19}
{Lustig-Yaeger}, J., {Meadows}, V.~S., \& {Lincowski}, A.~P. 2019, \aj, 158,
  27, \dodoi{10.3847/1538-3881/ab21e0}

\bibitem[{{Madden} \& {Kaltenegger}(2018)}]{madden18}
{Madden}, J.~H., \& {Kaltenegger}, L. 2018, Astrobiology, 18, 1559,
  \dodoi{10.1089/ast.2017.1763}

\bibitem[{{Malik} {et~al.}(2019){Malik}, {Kitzmann}, {Mendon{\c c}a}, {Grimm},
  {Marleau}, {Linder}, {Tsai}, \& {Heng}}]{malik19}
{Malik}, M., {Kitzmann}, D., {Mendon{\c c}a}, J.~M., {et~al.} 2019, arXiv
  e-prints.
\newblock \doarXiv{1903.06794}

\bibitem[{{Malik} {et~al.}(2017){Malik}, {Grosheintz}, {Mendon{\c c}a},
  {Grimm}, {Lavie}, {Kitzmann}, {Tsai}, {Burrows}, {Kreidberg}, {Bedell},
  {Bean}, {Stevenson}, \& {Heng}}]{malik17}
{Malik}, M., {Grosheintz}, L., {Mendon{\c c}a}, J.~M., {et~al.} 2017, \aj, 153,
  56, \dodoi{10.3847/1538-3881/153/2/56}

\bibitem[{{Mansfield} {et~al.}(2019){Mansfield}, {Kite}, {Hu}, {Koll}, {Malik},
  {Bean}, \& {Kempton}}]{mansfield19}
{Mansfield}, M., {Kite}, E.~S., {Hu}, R., {et~al.} 2019, arXiv e-prints,
  arXiv:1907.13150.
\newblock \doarXiv{1907.13150}

\bibitem[{Meadows {et~al.}(2018)Meadows, Arney, Schwieterman, Lustig-Yaeger,
  Lincowski, Robinson, Domagal-Goldman, Deitrick, Barnes, Fleming, Luger,
  Driscoll, Quinn, \& Crisp}]{meadows18}
Meadows, V.~S., Arney, G.~N., Schwieterman, E.~W., {et~al.} 2018, Astrobiology,
  18, 133, \dodoi{10.1089/ast.2016.1589}

\bibitem[{{Miller-Ricci} {et~al.}(2009){Miller-Ricci}, {Seager}, \&
  {Sasselov}}]{miller-ricci09}
{Miller-Ricci}, E., {Seager}, S., \& {Sasselov}, D. 2009, \apj, 690, 1056,
  \dodoi{10.1088/0004-637X/690/2/1056}

\bibitem[{{Morley} {et~al.}(2013){Morley}, {Fortney}, {Kempton}, {Marley},
  {Visscher}, \& {Zahnle}}]{morley13}
{Morley}, C.~V., {Fortney}, J.~J., {Kempton}, E.~M.-R., {et~al.} 2013, \apj,
  775, 33, \dodoi{10.1088/0004-637X/775/1/33}

\bibitem[{{Morley} {et~al.}(2015){Morley}, {Fortney}, {Marley}, {Zahnle},
  {Line}, {Kempton}, {Lewis}, \& {Cahoy}}]{morley15}
{Morley}, C.~V., {Fortney}, J.~J., {Marley}, M.~S., {et~al.} 2015, \apj, 815,
  110, \dodoi{10.1088/0004-637X/815/2/110}

\bibitem[{{Morley} {et~al.}(2017){Morley}, {Kreidberg}, {Rustamkulov},
  {Robinson}, \& {Fortney}}]{morley17}
{Morley}, C.~V., {Kreidberg}, L., {Rustamkulov}, Z., {Robinson}, T., \&
  {Fortney}, J.~J. 2017, \apj, 850, 121, \dodoi{10.3847/1538-4357/aa927b}

\bibitem[{Morris {et~al.}(2018)Morris, Agol, Davenport, \& Hawley}]{morris18}
Morris, B.~M., Agol, E., Davenport, J. R.~A., \& Hawley, S.~L. 2018, The
  Astrophysical Journal, 857, 39, \dodoi{10.3847/1538-4357/aab6a5}

\bibitem[{{Mulders} {et~al.}(2015){Mulders}, {Pascucci}, \& {Apai}}]{mulders15}
{Mulders}, G.~D., {Pascucci}, I., \& {Apai}, D. 2015, \apj, 814, 130,
  \dodoi{10.1088/0004-637X/814/2/130}

\bibitem[{Nickolls {et~al.}(2008)Nickolls, Buck, Garland, \& Skadron}]{cuda}
Nickolls, J., Buck, I., Garland, M., \& Skadron, K. 2008, Queue, 6, 40,
  \dodoi{10.1145/1365490.1365500}

\bibitem[{{Oliphant}(2007)}]{scipy}
{Oliphant}, T.~E. 2007, {Computing in Science \& Engineering}, 9,
  \dodoi{10.1109/MCSE.2007.58}

\bibitem[{{Owen} \& {Alvarez}(2016)}]{owen16}
{Owen}, J.~E., \& {Alvarez}, M.~A. 2016, \apj, 816, 34,
  \dodoi{10.3847/0004-637X/816/1/34}

\bibitem[{{Owen} \& {Wu}(2013)}]{owen13}
{Owen}, J.~E., \& {Wu}, Y. 2013, \apj, 775, 105,
  \dodoi{10.1088/0004-637X/775/2/105}

\bibitem[{{Owen} \& {Wu}(2017)}]{owen17}
---. 2017, \apj, 847, 29, \dodoi{10.3847/1538-4357/aa890a}

\bibitem[{{Pierrehumbert}(2010)}]{pierrehumbert10}
{Pierrehumbert}, R.~T. 2010, {Principles of Planetary Climate} (Cambridge
  University Press)

\bibitem[{{Pino} {et~al.}(2018{\natexlab{a}}){Pino}, {Ehrenreich},
  {Wyttenbach}, {Bourrier}, {Nascimbeni}, {Heng}, {Grimm}, {Lovis}, {Malik},
  {Pepe}, \& {Piotto}}]{pino18a}
{Pino}, L., {Ehrenreich}, D., {Wyttenbach}, A., {et~al.} 2018{\natexlab{a}},
  \aap, 612, A53, \dodoi{10.1051/0004-6361/201731244}

\bibitem[{{Pino} {et~al.}(2018{\natexlab{b}}){Pino}, {Ehrenreich}, {Allart},
  {Lovis}, {Brogi}, {Malik}, {Nascimbeni}, {Pepe}, \& {Piotto}}]{pino18b}
{Pino}, L., {Ehrenreich}, D., {Allart}, R., {et~al.} 2018{\natexlab{b}}, \aap,
  619, A3, \dodoi{10.1051/0004-6361/201832986}

\bibitem[{{Richard} {et~al.}(2012){Richard}, {Gordon}, {Rothman}, {Abel},
  {Frommhold}, {Gustafsson}, {Hartmann}, {Hermans}, {Lafferty}, {Orton},
  {Smith}, \& {Tran}}]{richard12}
{Richard}, C., {Gordon}, I.~E., {Rothman}, L.~S., {et~al.} 2012, \jqsrt, 113,
  1276, \dodoi{10.1016/j.jqsrt.2011.11.004}

\bibitem[{{Ricker} {et~al.}(2015){Ricker}, {Winn}, {Vanderspek}, {Latham},
  {Bakos}, {Bean}, {Berta-Thompson}, {Brown}, {Buchhave}, {Butler}, {Butler},
  {Chaplin}, {Charbonneau}, {Christensen-Dalsgaard}, {Clampin}, {Deming},
  {Doty}, {De Lee}, {Dressing}, {Dunham}, {Endl}, {Fressin}, {Ge}, {Henning},
  {Holman}, {Howard}, {Ida}, {Jenkins}, {Jernigan}, {Johnson}, {Kaltenegger},
  {Kawai}, {Kjeldsen}, {Laughlin}, {Levine}, {Lin}, {Lissauer}, {MacQueen},
  {Marcy}, {McCullough}, {Morton}, {Narita}, {Paegert}, {Palle}, {Pepe},
  {Pepper}, {Quirrenbach}, {Rinehart}, {Sasselov}, {Sato}, {Seager},
  {Sozzetti}, {Stassun}, {Sullivan}, {Szentgyorgyi}, {Torres}, {Udry}, \&
  {Villasenor}}]{ricker15}
{Ricker}, G.~R., {Winn}, J.~N., {Vanderspek}, R., {et~al.} 2015, Journal of
  Astronomical Telescopes, Instruments, and Systems, 1, 014003,
  \dodoi{10.1117/1.JATIS.1.1.014003}

\bibitem[{{Rogers} \& {Seager}(2010)}]{rogers10}
{Rogers}, L.~A., \& {Seager}, S. 2010, \apj, 712, 974,
  \dodoi{10.1088/0004-637X/712/2/974}

\bibitem[{{Rothman} {et~al.}(2010){Rothman}, {Gordon}, {Barber}, {Dothe},
  {Gamache}, {Goldman}, {Perevalov}, {Tashkun}, \& {Tennyson}}]{rothman10}
{Rothman}, L.~S., {Gordon}, I.~E., {Barber}, R.~J., {et~al.} 2010, \jqsrt, 111,
  2139, \dodoi{10.1016/j.jqsrt.2010.05.001}

\bibitem[{{S{\'a}nchez-Lavega} {et~al.}(2004){S{\'a}nchez-Lavega},
  {P{\'e}rez-Hoyos}, \& {Hueso}}]{sanchez-lavega04}
{S{\'a}nchez-Lavega}, A., {P{\'e}rez-Hoyos}, S., \& {Hueso}, R. 2004, American
  Journal of Physics, 72, 767, \dodoi{10.1119/1.1645279}

\bibitem[{{Schaefer} {et~al.}(2016){Schaefer}, {Wordsworth}, {Berta-Thompson},
  \& {Sasselov}}]{schaefer16}
{Schaefer}, L., {Wordsworth}, R.~D., {Berta-Thompson}, Z., \& {Sasselov}, D.
  2016, \apj, 829, 63, \dodoi{10.3847/0004-637X/829/2/63}

\bibitem[{{Seemann} {et~al.}(2014){Seemann}, {Anglada-Escude}, {Baade},
  {Bristow}, {Dorn}, {Follert}, {Gojak}, {Grunhut}, {Hatzes}, {Heiter}, {Ives},
  {Jeep}, {Jung}, {K{\"a}ufl}, {Kerber}, {Klein}, {Lizon}, {Lockhart},
  {L{\"o}winger}, {Marquart}, {Oliva}, {Paufique}, {Piskunov}, {Pozna},
  {Reiners}, {Smette}, {Smoker}, {Stempels}, \& {Valenti}}]{seemann14}
{Seemann}, U., {Anglada-Escude}, G., {Baade}, D., {et~al.} 2014, in \procspie,
  Vol. 9147, Ground-based and Airborne Instrumentation for Astronomy V, 91475G

\bibitem[{{Selsis} {et~al.}(2007){Selsis}, {Chazelas}, {Bord{\'e}}, {Ollivier},
  {Brachet}, {Decaudin}, {Bouchy}, {Ehrenreich}, {Grie{\ss}meier}, {Lammer},
  {Sotin}, {Grasset}, {Moutou}, {Barge}, {Deleuil}, {Mawet}, {Despois},
  {Kasting}, \& {L{\'e}ger}}]{selsis07}
{Selsis}, F., {Chazelas}, B., {Bord{\'e}}, P., {et~al.} 2007, \icarus, 191,
  453, \dodoi{10.1016/j.icarus.2007.04.010}

\bibitem[{{Sneep} \& {Ubachs}(2005)}]{sneep05}
{Sneep}, M., \& {Ubachs}, W. 2005, \jqsrt, 92, 293

\bibitem[{{Snellen} {et~al.}(2010){Snellen}, {de Kok}, {de Mooij}, \&
  {Albrecht}}]{snellen10}
{Snellen}, I.~A.~G., {de Kok}, R.~J., {de Mooij}, E.~J.~W., \& {Albrecht}, S.
  2010, \nat, 465, 1049, \dodoi{10.1038/nature09111}

\bibitem[{Sousa-Silva {et~al.}(2015)Sousa-Silva, Al-Refaie, Tennyson, \&
  Yurchenko}]{sousasilva15}
Sousa-Silva, C., Al-Refaie, A.~F., Tennyson, J., \& Yurchenko, S.~N. 2015,
  Monthly Notices of the Royal Astronomical Society, 446, 2337,
  \dodoi{10.1093/mnras/stu2246}

\bibitem[{{Southworth} {et~al.}(2017){Southworth}, {Mancini}, {Madhusudhan},
  {Molli{\`e}re}, {Ciceri}, \& {Henning}}]{southworth17}
{Southworth}, J., {Mancini}, L., {Madhusudhan}, N., {et~al.} 2017, \aj, 153,
  191, \dodoi{10.3847/1538-3881/aa6477}

\bibitem[{{Spiegel} \& {Burrows}(2010)}]{spiegel10}
{Spiegel}, D.~S., \& {Burrows}, A. 2010, \apj, 722, 871,
  \dodoi{10.1088/0004-637X/722/1/871}

\bibitem[{{Stock} {et~al.}(2018){Stock}, {Kitzmann}, {Patzer}, \&
  {Sedlmayr}}]{stock18}
{Stock}, J.~W., {Kitzmann}, D., {Patzer}, A.~B.~C., \& {Sedlmayr}, E. 2018,
  \mnras, 479, 865, \dodoi{10.1093/mnras/sty1531}

\bibitem[{Thalman {et~al.}(2014)Thalman, J.~Zarzana, Tolbert, \&
  Volkamer}]{thalman14}
Thalman, R., J.~Zarzana, K., Tolbert, M., \& Volkamer, R. 2014, Journal of
  Quantitative Spectroscopy and Radiative Transfer, 147, 171–177,
  \dodoi{10.1016/j.jqsrt.2014.05.030}

\bibitem[{{Tian}(2009)}]{tian09}
{Tian}, F. 2009, \apj, 703, 905, \dodoi{10.1088/0004-637X/703/1/905}

\bibitem[{{van der Walt} {et~al.}(2011){van der Walt}, {Colbert}, \&
  {Varoquaux}}]{numpy}
{van der Walt}, S., {Colbert}, S.~C., \& {Varoquaux}, G. 2011, Computing in
  Science and Engineering, 13, 22, \dodoi{10.1109/MCSE.2011.37}

\bibitem[{{van Rossum}(1995)}]{python}
{van Rossum}, G. 1995, {Python} tutorial, Technical Report CS-R9526, Centrum
  voor Wiskunde en Informatica (CWI), Amsterdam

\bibitem[{{Vanderspek} {et~al.}(2019){Vanderspek}, {Huang}, {Vanderburg},
  {Ricker}, {Latham}, {Seager}, {Winn}, {Jenkins}, {Burt}, {Dittmann},
  {Newton}, {Quinn}, {Shporer}, {Charbonneau}, {Irwin}, {Ment}, {Winters},
  {Collins}, {Evans}, {Gan}, {Hart}, {Jensen}, {Kielkopf}, {Mao}, {Waalkes},
  {Bouchy}, {Marmier}, {Nielsen}, {Ottoni}, {Pepe}, {S{\'e}gransan}, {Udry},
  {Henry}, {Paredes}, {James}, {Hinojosa}, {Silverstein}, {Palle},
  {Berta-Thompson}, {Crossfield}, {Davies}, {Dragomir}, {Fausnaugh}, {Glidden},
  {Pepper}, {Morgan}, {Rose}, {Twicken}, {Villase{\~n}or}, {Yu}, {Bakos},
  {Bean}, {Buchhave}, {Christensen-Dalsgaard}, {Christiansen}, {Ciardi},
  {Clampin}, {De Lee}, {Deming}, {Doty}, {Jernigan}, {Kaltenegger}, {Lissauer},
  {McCullough}, {Narita}, {Paegert}, {Pal}, {Rinehart}, {Sasselov}, {Sato},
  {Sozzetti}, {Stassun}, \& {Torres}}]{vanderspek19}
{Vanderspek}, R., {Huang}, C.~X., {Vanderburg}, A., {et~al.} 2019, \apjl, 871,
  L24, \dodoi{10.3847/2041-8213/aafb7a}

\bibitem[{Wagner \& Kretzschmar(2008)}]{wagner08}
Wagner, W., \& Kretzschmar, H.-J. 2008, International Steam Tables - Properties
  of Water and Steam Based on the Industrial Formulation IAPWS-IF97 No.
  978-3-540-74234-0 (Springer, Berlin, Heidelberg),
  \dodoi{https://doi.org/10.1007/978-3-540-74234-0}

\bibitem[{{Wolf}(2017)}]{wolf17}
{Wolf}, E.~T. 2017, \apjl, 839, L1, \dodoi{10.3847/2041-8213/aa693a}

\bibitem[{{Yurchenko} {et~al.}(2011){Yurchenko}, {Barber}, \&
  {Tennyson}}]{yurchenko11}
{Yurchenko}, S.~N., {Barber}, R.~J., \& {Tennyson}, J. 2011, \mnras, 413, 1828,
  \dodoi{10.1111/j.1365-2966.2011.18261.x}

\bibitem[{{Yurchenko} \& {Tennyson}(2014)}]{yurchenko14}
{Yurchenko}, S.~N., \& {Tennyson}, J. 2014, \mnras, 440, 1649,
  \dodoi{10.1093/mnras/stu326}

\bibitem[{{Zhang} {et~al.}(2018){Zhang}, {Zhou}, {Rackham}, \&
  {Apai}}]{zhang18}
{Zhang}, Z., {Zhou}, Y., {Rackham}, B.~V., \& {Apai}, D. 2018, \aj, 156, 178,
  \dodoi{10.3847/1538-3881/aade4f}

\end{thebibliography}

\end{document}